\documentclass[fleqn,usenatbib]{mnras}

\usepackage{newtxtext,newtxmath}
\usepackage[T1]{fontenc}

\DeclareRobustCommand{\VAN}[3]{#2}
\let\VANthebibliography\thebibliography
\def\thebibliography{\DeclareRobustCommand{\VAN}[3]{##3}\VANthebibliography}

\usepackage{graphicx}	% Including figure files
\usepackage{amsmath}	% Advanced maths commands
\usepackage{booktabs}
\usepackage{pdflscape}
\newcommand{\msun}{M_{\odot}}

\title[Recovering orbits of satellites]{Uncertainties associated with the backward integration of dwarf satellites using simple parametric potentials}

\author[D'Souza \& Bell]{Richard D'Souza$^{1}$ 
\thanks{Contact e-mail:\href{mailto:rdsouza@speola.va}{rdsouza@specola.va}},
						Eric F.\ Bell$^{2}$\\
$^{1}$Vatican Observatory, Specola Vaticana, V-00120, Vatican City State\\
$^{2}$University of Michigan, Department of Astronomy, 311 West Hall, 1085 South University Ave., Ann Arbor, MI 48109-1107
}

\date{Accepted XXX. Received YYY; in original form ZZZ}

\pubyear{2022}

\begin{document}
\label{firstpage}
\pagerange{\pageref{firstpage}--\pageref{lastpage}}
\maketitle

\begin{abstract}
In order to backward integrate the orbits of Milky Way (MW) dwarf galaxies, much effort has been invested in recent years to constrain their initial phase-space coordinates. Yet equally important are the assumptions on the potential that the dwarf galaxies experience over time, especially given the fact that the MW is currently accreting the Large Magellanic Cloud (LMC). In this work, using a dark matter-only zoom-in simulation, we test whether the use of common parametric forms of the potential is adequate to successfully backward integrate the orbits of the subhaloes from their present-day positions. We parametrise the recovered orbits and compare them with those from the simulations. We find that simple symmetric parametric forms of the potential fail to capture the complexities and the inhomogeneities of the true potential experienced by the subhaloes. More specifically, modelling a recent massive accretion like that of the LMC as a sum of two spherical parametric potentials leads to substantial errors in the recovered parameters of the orbits. These errors rival those caused due to a) a 30\% uncertainty in the virial mass of the MW and b) not modelling the potential of the recently accreted massive satellite. Our work suggests that i) the uncertainties in the parameters of the recovered orbits of some MW dwarfs may be under-estimated and that ii) researchers should characterise the uncertainties inherent to their choice of integration techniques and assumptions of the potential against cosmological zoom-in simulations of the MW, which include a recently-accreted LMC.
\end{abstract}


%%%%%%%%%%%%%%%%%%%%%%%%%%%%%%%%%%%%%%%%%%%%%%%%%%

%%%%%%%%%%%%%%%%% BODY OF PAPER %%%%%%%%%%%%%%%%%%

\section{Introduction}
Accurate backward integration of the orbits are central to answer a number of pressing questions about the population of Milky Way (MW) dwarf galaxies. For example, the number of dwarf galaxies contributed by the Large Magellanic Cloud (LMC) to the Galaxy is a question of much interest. While a number of ultra-faint dwarf galaxies have been discovered in close proximity to the LMC \citep[e.g.][]{Kallivayalil2018}, it remains a matter of debate as to which MW classical dwarf spheroidals were once satellites of the LMC (e.g. Fornax and Carina, \citealt{Pardy2020,Erkal2019}, although see \citealt{Patel2020} for a different view). Similarly, attempts are being made to understand which of the MW's satellites and globular clusters were accreted along with its ancient massive accretions \citep[e.g.][]{Boldrini2021}. Such backward integrations are also used to try to understand the impacts of interactions on the properties or star formation histories of galaxies. For example, improved constraints on the orbit of the Sagittarius dwarf galaxy can help us to better understand its influence on the stellar disk \citep[e.g.][]{Gomez2013,Laporte2018,Ruiz-Lara2020}, while the bursts of star formation in the Fornax dwarf spheroidal galaxy are though to be associated with its pericentric passages \citep{Rusakov2021}. Retracing orbital histories to constrain the infall time of satellites can give us an insight into the quenching of star formation in the MW dwarf spheroidals \citep[e.g.][]{Weisz2015,Fillingham2019,Rusakov2021}. The accuracy with which we can answer these and other questions about the MW dwarf galaxy population depends in great part on our ability to accurately constrain their orbital histories.

Our ability to backward integrate the orbit of a MW dwarf galaxy rests on two equally important ingredients --- {\it i)} knowledge of the dwarf galaxy's present-day position and motions and {\it ii)} the potential of the MW and its evolution over time. In recent years, our observational constraints on the full 6-D space coordinates of the MW dwarf galaxies have improved by leaps and bounds. Distance estimates to the classical MW dwarf galaxies can now be estimated to better than 5\% using RR Lyrae stars while the distances to smaller dwarf galaxies can be estimated to better than 10\% precision \citep[e.g.][]{Hernitschek2019}, with further improvements expected with the upcoming Rubin observatory \citep{Oluseyi2012}. On the other hand, recent data from HST and Gaia have revolutionised our constraints on the proper motions of the MW dwarf galaxies \citep[e.g.][]{Fritz2018,Simon2018,McConnachie2020,Battaglia2021}. Future Gaia data releases are expected to reduce the uncertainty in the measured proper motions by a further factor of 4 or more. While our constraints on the initial conditions of the MW dwarf galaxies will only get better with time, our ability to backward integrate their orbits is limited by the uncertainty in the constraints on the potential of the galaxy and its evolution over time.

Our knowledge about the present-day potential of the MW is far from perfect. The primary uncertainty of the potential is the total mass of the MW and its mass distribution. Estimates of the mass of the MW range from $8\times10^{11}\msun$ to $1.5\times10^{12}\msun$ \citep[$\sim$30\% uncertainty in the mass of the MW, e.g.][]{Eadie2019,Callingham2019,Cautun2020,Deason2021,Magnus2021}. While the mass-concentration relationship \citep[e.g.][]{Bullock2001,Wechsler2002} resulting from the hierarchical assembly of haloes provides a strong prior for the concentration of the MW, its mass distribution can be measured using tracers at various radial distances. Thanks to recent data, our constraints on the mass of the halo in the inner part of the Galaxy have improved considerably. For example, Gaia proper motion data of standard candles in combination with data from other spectroscopic surveys have allowed us to tightly constrain the amplitude and shape of the potential in the inner part of the Galaxy dominated by its stellar disk \citep{Nitschai2020,Nitschai2021}. Further out, one has to rely on a number of diverse and scarce dynamical tracers including the globular cluster system \citep[e.g.,][]{Watkins2019,Posti2019}, stellar streams \citep[e.g.][]{Bovy2016}, distant halo stars \citep{Deason2021} as well as the satellite galaxy population \citep[e.g.,][]{Fritz2020}. While some of these more distant tracers allow one to constrain the mass profile of the galaxy out to 100 kpc, the lack of complete dynamical equilibrium in the outer tracer population due to the recent accretion of the LMC imposes systematic uncertainties in their masses \citep[25-35\%;][]{Fritz2020,Erkal2020b}. In the absence of more distant tracers, one is forced to use models to extrapolate the mass of the galaxy out to the virial radius, thereby introducing further systematic uncertainty in the outer potential of the Galaxy.

A further source of uncertainty is the shape of the galaxy. While the DM halo of the MW is expected to be triaxial, baryonic effects are thought to make the DM halo more spherical in its inner parts \citep[][]{Bryan2013}. Various methods have been proposed to measure the shape of the inner DM halo. The kinematics and the spatial distribution of thin stellar streams can be used to measure shape of the potential \citep[e.g.][]{Johnston2005,Law2010,Deg2013,Price-Whelan2014,Bovy2016}. The shape of the halo can also be inferred from the kinematics of globular clusters or halo field stars while imposing the condition that these tracers are in dynamical equilibrium \citep[e.g.][]{Posti2019,Hattori2021}. Furthermore, hypervelocity stars originating from the galactic center can also allow us to probe the shape of the DM matter halo \citep[e.g.][]{Gnedin2005,Contigiani2019}. Using the globular cluster system, \cite{Posti2019} infer that the inner 20 kpc of the MW is strongly prolate, a result which also agrees with the results from a single hypervelocity star \citep{Hattori2020}. Yet, no constraints on the shape of the halo exists for a large part of the radial range over which we backward integrate the orbits of dwarf galaxies.

The growth history of the MW in the last 7 Gyr is also uncertain. Cosmological simulations inform us that the average mass of a MW-mass halo grows appreciably in the time a dwarf satellite galaxy takes to complete a single orbit \citep[e.g.][]{Diemer2013}. The accretion of the Magellanic Clouds (likely to be at least 1/10 of the mass of the MW, and perhaps as much as 1/4; \citealt{Gomez2015,Penarrubia2016}) in the last few Gyr is expected to have contributed substantially to the mass growth of the MW. Similar periods of growth history are expected with the accretion of the Gaia Enceladus Sausage galaxy \citep{Belokurov2018,Helmi2018}. In principle, measuring the dispersion of thin early accreted Galactic streams can give us constraints on the growth history of the MW\citep{Buist2015}. However, this technique has yet to be exploited to get any useful constraints. The uncertainty in the evolution of the potential of the MW presents substantial difficulties in our ability to recover the orbital histories of the MW satellite galaxies.

Finally, another source of uncertainty in backward integrating the orbits is introduced by the use of simple parametric forms to describe the complex 3D potential of the Galaxy. In general, the total potential of the Galaxy is described by a sum of parametric models describing the bulge, the disk and the dark matter halo respectively, where the profile of the halo has been informed by cold dark matter-only cosmological simulations \citep[e.g., NFW;][]{NFW1997}. While these models have the advantages of being described by a few parameters, they often fail to describe the complex 3D shape of a real halo. In the absence of constraints on the 3D shape of the halo especially in the outer parts, simple spherical models are often used.  Insights from hydrodynamical simulations (which attempt to model baryons and dark matter self-consistently) suggest that the baryons cause a contraction of the dark matter halo, thereby introducing a systemic bias in the assumed potential \citep[e.g. see][]{Cautun2020}. Moreover, the smooth functional form of the models fails to reproduce the small-scale perturbations in the halo due to presence of substructure. Finally, the LMC is very likely to be on it first infall \citep[e.g.][]{Besla2007}, just past its first pericentric passage, causing large time-dependent structural changes to the halo on shorter time frames \citep{Wang2020} including a wake \citep{Garavito-Camargo2019, Conroy2021} and a distance-dependent clustering in the orbital poles of the satellites \citep{Garavito-Camargo2021b}. Simple parametric models fail to capture these strong temporary inhomogeneities introduced in the potential of the halo.

A number of attempts have been made in the literature to backward integrate the orbits of the MW dwarf galaxies --- ranging from a straight-forward backward integration in a static MW potential to increasingly sophisticated modelling of the potential in recent years, taking into account the contribution of the infalling LMC and SMC \citep[][]{Patel2020,Vasiliev2021}. Not only does the LMC affect the present-day orbit of the Orphan stream \citep{Erkal2019}, the Sagittarius stream \citep{Vasiliev2021} as well as 5 streams in the Southern Galactic Hemisphere \citep{Shipp2021}, but it has also been demonstrated that the LMC dramatically affects the orbits of a number of MW dwarf galaxies \citep[e.g.][]{Patel2020,Battaglia2021}. Approximate techniques have also been developed to constrain the infall time of the MW dwarf galaxies \citep[e.g.][]{Weisz2015,Fillingham2019}, taking advantage of the correlation between the present day orbital energy of a dwarf galaxy and its infall time for halos that grow slowly \citep{Rocha2012}. Yet all these attempts, even the most sophisticated of them, use parametric models to describe the potential of the MW halo. While attempts have been made to characterise the errors due to uncertainties in the proper motion and distances as well as uncertainties in the total mass of the MW and the LMC, we still do not understand the uncertainties in backward integrating the orbits of MW dwarf galaxies arising from the use of the simple parametric forms to characterise the potential.

Recently a few attempts have been made to move beyond simple parametric models and characterise through basis function expansions the evolution of a simulated isolated MW-mass halo \citep{Sanders2020} or a MW-mass halo which suffered a recent massive accretion \citep{Cunningham2020,Garavito-Camargo2021}. \cite{Sanders2020} have demonstrated that they can achieve high fidelity	orbit reconstruction of the orbits of subhaloes in the case of an isolated MW-mass halo using at least 15 radial and 6 angular terms in the basis expansion and a $\sim100$Myr temporal resolution. Since constraining the coefficients of the basis functions of a real life situation like that of the MW and the LMC is an ardous task, most works resort to simple parameteric models of the potential for their convenience.

In this work, we make a first attempt to study how the use of simple parametric form to model the potential affects our reconstruction of the orbital histories of MW satellites. We do this by considering the simplest possible case: i.e., a dark matter-only zoom-in simulation of a MW-mass halo. Given perfect knowledge(!) about the 6-d phase space coordinates of the subhaloes of the MW-mass halo at $\mathrm{z=0}$ and an assumed potential, we can backward integrate their orbits and compare them with the real orbit from the simulations. In backward integrating the orbits, we assume that the subhalo experiences only the potential of the central MW-mass host and recently accreted LMC-like massive satellite. We approximate the potential of the central MW-like host and the LMC-like massive accreted satellite using simple parametric models --- whose respective masses and concentrations are taken from the merger trees of the simulation. While these parameteric models may be relatively simple, they correspond to the state-of-the-art techniques used for inferring the orbits of the MW satellites \citep{Patel2020,Battaglia2021}. Furthermore, we parametrise the recovered orbits of the subhaloes in a number of ways (e.g., the distance of the first and second pericentre/apocentres or the time of accretion onto the halo) depending on several scientific questions in the literature. By comparing the parameters of the recovered orbit with those of the true orbit, we can study how the use of simple parametric models affects our recovery of those parameters.

The plan of this paper is as follows. We describe the simulations, our assumptions and the numerical techniques in Section \ref{sec:methods}, and the ways we choose to parametrise the orbit of a subhalo. In Section \ref{sec:energetics}, we examine the relation between the present-day binding energies of the subhaloes and their time of accretion. In Section \ref{sec:backintegrate}, we demonstrate how well we can recover the parameters of the orbits of the subhaloes in a MW-mass halo which has not suffered a massive accretion, and in one which has suffered a massive accretion, respectively. In Section \ref{sec:compare}, we characterise the errors due to the use of parametric models of the potential with respect to the other sources of uncertainty. In Section \ref{sec:conc}, we summarise and discuss our findings.

\section{Numerical Methods}
\label{sec:methods}
\subsection{Simulations}
We use a suite of 48 high-resolution, zoom-in dark matter-only simulations of MW-mass haloes \citep[ELVIS simulation;][]{Garrison2014} with virial masses ranging from $\mathrm{M_{vir}} = 1-3 \times 10^{12}\, \mathrm{\msun}$. The particle resolution is $1.9 \times 10^5 \,\mathrm{\msun}$ and the Plummer-equivalent force softening is 140 pc. ELVIS assumes a Wilkinson Microwave Anisotropy Probe cosmology \citep[WMAP7,][$\Omega_m=0.286$, $\Omega_\Lambda=0.714$, $h=0.7$, $\sigma_8=0.82$, and $n_s=0.96$]{Larson2011}. Each simulation contains a maximum of 75 snapshots with the average temporal spacing at about 250 Myr. Dark matter subhaloes are identified using ROCKSTAR \citep{Rockstar}, while merger trees are constructed using CONSISTENT-TREES algorithm \citep{Consistent_trees}. The virial masses of the subhaloes ($\mathrm{M_{vir}}$)	 are reported as the mass within a sphere of radius $\mathrm{R_{vir}}$ that corresponds to an overdensity of 97 relative to the critical density of the Universe \citep{Byran1998}. For each subhalo, its primary progenitor (main branch) is identified as the progenitor that contains the largest total mass summed from the subhalo masses over all preceding snapshots in that branch. The main progenitor branch of each host MW-mass halo is isolated through the primary progenitors. The peak mass ($\mathrm{M_{peak}}$) of each subhalo is the maximum mass the subhalo has had over its entire main progenitor branch. Further detail of the ELVIS simulations can be found in \cite{Garrison2014}.

We restrict our attention to subhaloes with $\mathrm{M_{peak}} > 10^9\, \mathrm{\msun}$ which we assume are capable of hosting dwarf galaxies, some of them being classical dwarf satellites and which are not susceptible to artificial disruption \citep[e.g.][]{vandenBosch2018a, vandenBosch2018b}. We calculate the time of accretion ($\mathrm{T_{Acc}}$) of each subhalo as the time when the subhalo first crosses the virial radius of the main subhalo. In order to explore the the impact of massive mergers (such as the LMC and Gaia-Enceladus) on the backward reconstruction of satellite orbits, we identify {\it massive progenitors} (merged) and {\it satellites} (surviving) accreted onto the MW-mass host to be those whose peak mass is greater than 10\% of the present day virial mass of the MW-mass halo \citep{DSouza2021}.

\begin{figure}
	\includegraphics[trim=0 20 0 0, clip,width=\columnwidth]{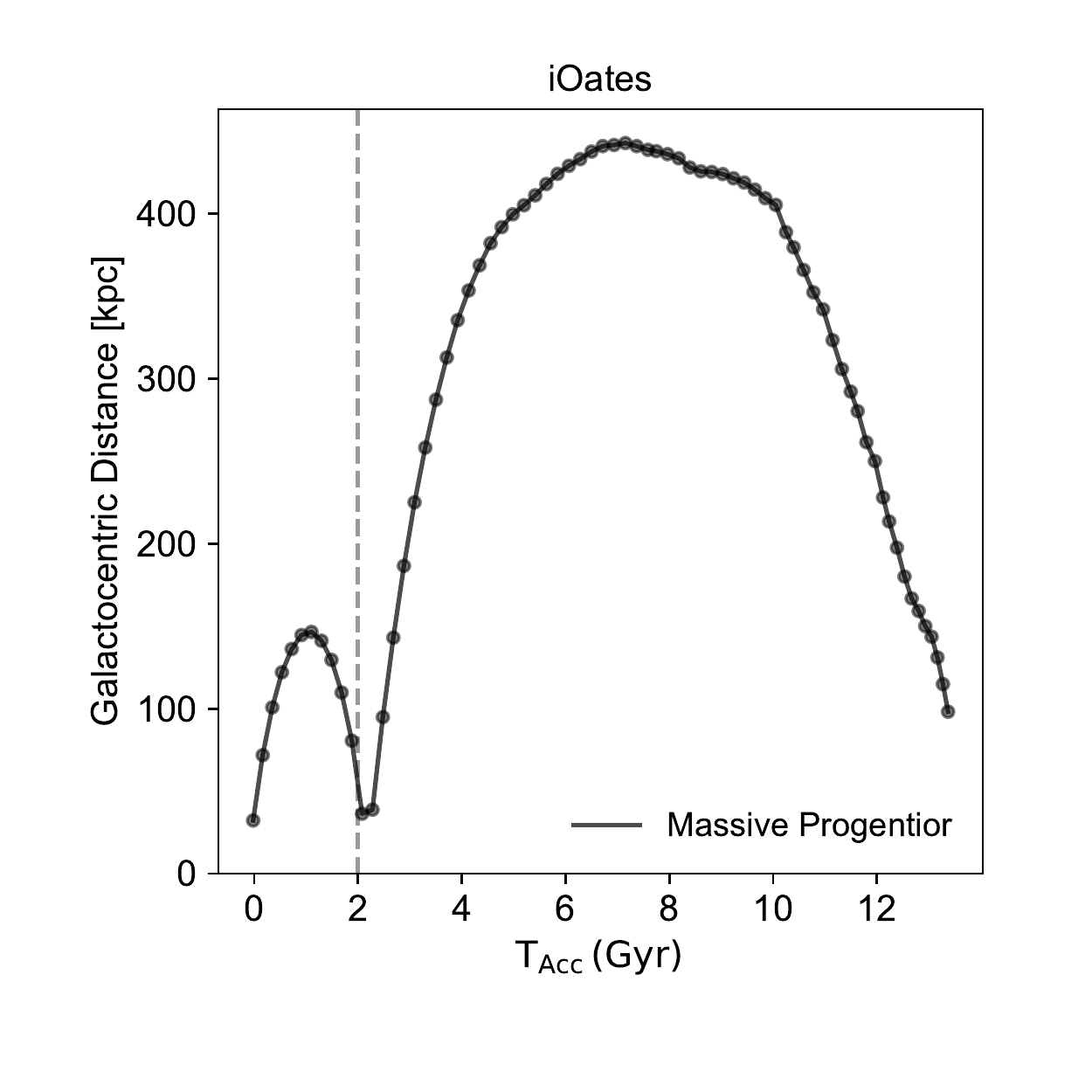}
	\caption{The MW-mass halo `iOates' taken from the ELVIS suite of simulations best represents the MW. It recently accreted a massive satellite about the mass of the LMC (~1:10 merger). We present the galactocentric distance of the massive satellite as a function of the original lookback time for the MW-mass halo `iOates'. The vertical dashed line represents the time which best reproduces the current distance between the MW and the LMC. Hence, for `iOates', we start our backward integration of the subhaloes from this particular point, and we redefine the time to accretion of the subhaloes.}
	\label{fig:fig1}
\end{figure}

We illustrate our main results using two particular MW-mass haloes: `iDouglas', a MW-mass halo which did not accrete any massive progenitor or massive satellite ($>10^{11}\,\msun$) during its lifetime and `iOates', which recent accreted a massive satellite. While `iDouglas' is a prototype of a MW-mass halo which grew in isolation, `iOates' represents a halo similar to the MW and the LMC. We note that although `iOates' suffered a $\sim$1:10 merger, relatively few subhaloes are accreted along with its recent massive satellite. In Fig. \ref{fig:fig1}, we plot the galactocentric distance of the massive satellite of `iOates' as a function of time. The first pericentric passage is at a distance of roughly 37 kpc about $\sim2$ Gyr ago.  In order to better represent the case of the MW and the LMC, we start the backward integration from the point just after the first pericentre as represented by the vertical dashed line in Fig. \ref{fig:fig1}, and redefine the time of accretion of all the subhaloes with reference to this point of time.

\subsection{Backward integration}
We attempt to backward integrate the orbits of the subhaloes of a MW-mass host, using a modified version of \texttt{Galpy} \citep[][http://github.com/jobovy/galpy]{Bovy2015} and their current phase-space coordinates as initial conditions. In backward integrating the orbits of a subhalo, we assume that:
\begin{itemize}
	\item A subhalo experiences only the potential due to the MW-mass host and the most recent massive accreted satellite. The impacts on the potential from any other massive accretions are not modelled. This implicitly assumes that subhaloes are accreted in one of two ways --- either onto the MW-mass host individually or accompanying the most recent massive accreted satellite.
	\item The total potential experienced by the subhalo is the sum of the potential of the MW-mass host and the most recent massive satellite, which are modelled using spherical parametric forms. The MW-mass host is modelled with a spherical NFW profile \citep{NFW1997}, while the massive accreted satellite is modelled as a spherical Hernquist profile \citep{Hernquist1990}. The choice of the parametric forms of the potential (common in the literature; see \citealt{Patel2020}) ensures that the mass of the massive satellite is finite and that an isolated subhalo at large galactocentric distances experiences primarily the potential of the MW-mass host. Triaxiality or inhomogeneities in the total potential are neglected.
\end{itemize}

We include the true mass-growth of the of the central MW-mass host as well as the most recent massive satellite to calculate the potential, while considering the effect of keeping the mass of the MW fixed over time in Appendix \ref{appendix:sec2}. Using the masses, radii and concentration of the MW-mass host and the massive satellite recorded in the ROCKSTAR catalogue and the CONSISTENT merger-trees, we implement the mass growth using a step-function, i.e., we update the potential at the beginning of each snapshot, while allowing it to remain constant between snapshots. This compromise allows us to use fast-integrators of Galpy while allowing us to implement the mass growth of the MW-mass host. In Appendix \ref{appendix:timestep}, we estimate that this choice of mass-step contributes less than 10\% to the total error budget.

We backward integrate the orbits of only the subhaloes capable of hosting dwarf galaxies, considering them as test particles experiencing only the potential due to the central MW-mass halo and its recently accreted massive satellite (a restricted three body problem). Note that we do not backward integrate the path of the recently accreted massive satellite; instead, we take its path directly from the N-body model. In addition, the subhaloes also experience various inertial forces in the reference frame of the MW due to its reflex motion caused by the recently accreted massive satellite as well as the motion of the MW-mass host through the large scale structure over cosmic time. We account for these inertial forces in the following way. We model the central MW-mass halo and its recently accreted massive satellite as moving potentials whose phase space coordinates are taken directly from the merger trees. Furthermore, we calculate the orbits in a moving reference frame, which we estimate by fitting a fourth-order polynomial to the motion of the MW-mass host. This ensures that the centre of the MW-mass host is never more than 60 kpc away from the centre of the reference frame, allowing for great accuracy in our backward integrations in the adopted cylindrical coordinate system. We account for the fictious forces arising from our choice of the non-inertial reference frame \citep{Sanders2020}. That is the total force on a test particle is 
\begin{equation}
	\mathbf{F}(\mathbf{x},t) =  - \nabla \Phi (\mathbf{x},t) - \dot{u}(t)
\end{equation}
where $\Phi (\mathbf{x},t)$ is the halo potential and $u(t)$ is the pecuilar velocity of the reference frame.

We include approximate prescriptions for accounting for the dynamical friction on the subhaloes from both the MW-mass halo and its most-recent massive accretion. To account for the dynamical friction due to the central MW-mass halo, we use \texttt{Galpy}'s standard dynamical friction prescription --- a model based on the Chandrasekhar formula \citep{Chandrasekhar1943} that agrees well with $N$-body results for infalling cored and cusped haloes --- for the central MW-mass halo \citep{Petts2016}. In modelling the typically much smaller dynamical friction on subhaloes from the massive accreted satellite, we choose to follow \cite{Patel2020} in adopting a simpler approach:
\begin{equation}
	\mathrm{\mathbf{F}}_{\mathrm{PS},DF}=0.428 \ln \Lambda \frac{\mathrm{G\,M^{2}_{sub}}}{\mathrm{r_{PS}}^2} \frac{\mathbf{v}_{\mathrm{PS}}}{v_{\mathrm{PS}}}
\end{equation}
with the Coulomb logarithm as $\ln \Lambda=0.3$ and $\mathrm{r_{PS}}$ and $v_{\mathrm{PS}}$ are the distance and the relative velocity between the subhalo and the massive accreted satellite. The dynamical friction experienced by the subhaloes due to the massive accreted satellite is much less than the dynamical friction from the central MW-mass host.

For our backward integrations, we use 25 timesteps between each snapshot, i.e., an average timestep of 10 Myr. Tests with increasing the number of timesteps demonstrate that the orbits are stable and that the integrations have converged.

We also calculate the binding energy (BE=PE-KE) and the circularity parameter \citep[$\epsilon_{J}=J_z/J_{\mathrm{circ}}(E)$;][]{Abadi2003} of each subhalo at $\mathrm{z=0}$.

\subsection{Parametrising the orbit}
\label{subsec:parameter}
\begin{figure}
	\begin{center}
	\includegraphics[trim=0 40 0 0, clip,width=\columnwidth]{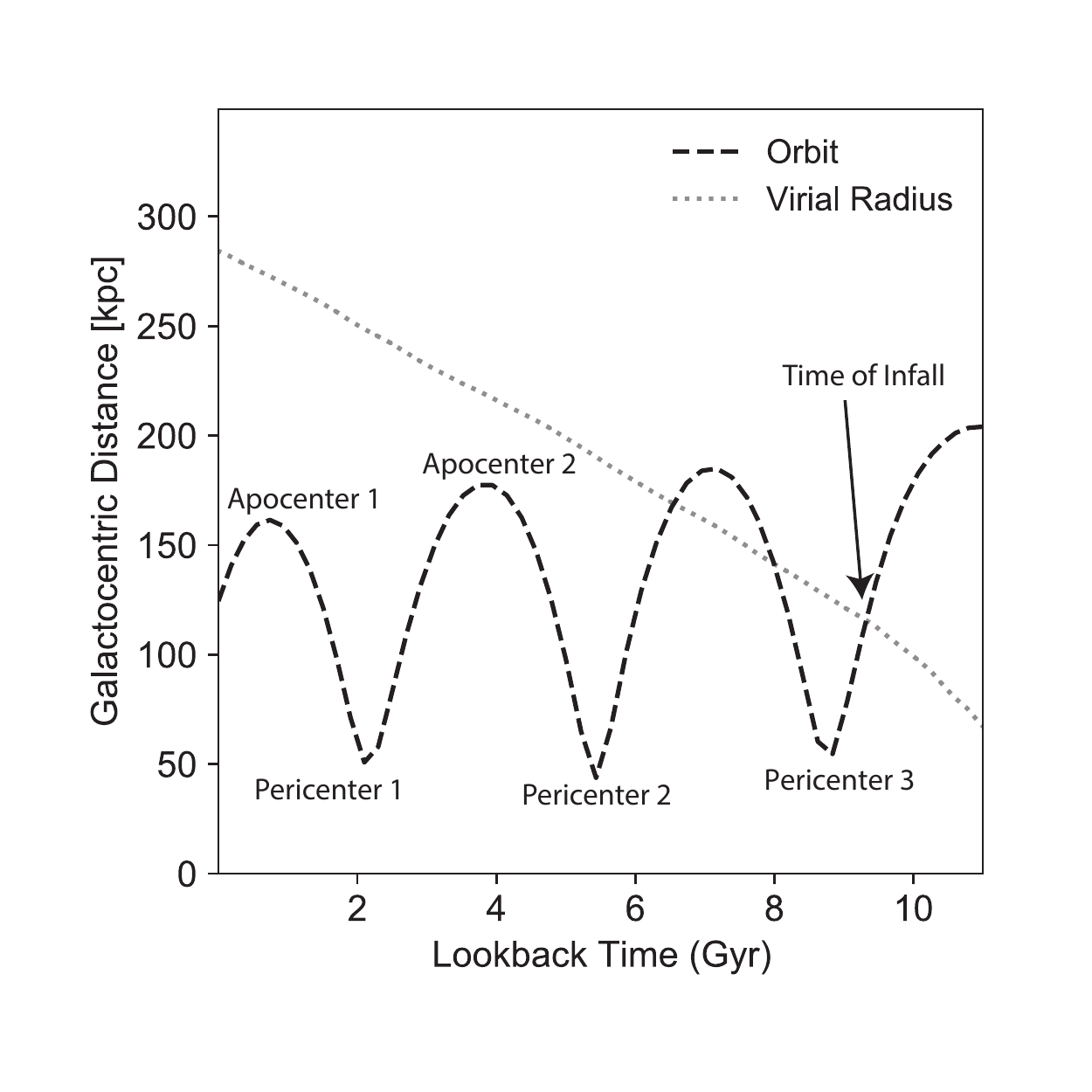}
	\caption{Parametrization of the orbits of the subhaloes based on several relevant science cases. Using a randomly chosen orbit, we graphically indicate the main parameters we choose to measure from the real and the backward integrated orbits of the subhaloes (elucidated in Table \ref{table:table1}) in terms of distance reached (in kpc) and lookback time (in Gyr).}
	\label{fig:fig2}
	\end{center}
\end{figure}

In this work, we choose to parameterise the orbits of the subhaloes based on the three broad science cases. First, to understand star formation bursts in large MW dwarf spheroidals \citep{Rusakov2021} as well as the effect of the Sagittarius dwarf galaxy on the MW disk \citep{Laporte2018}, we identify the time and the galactocentric distance of the most recent pericentric passages with respect to the central MW-mass host \citep[e.g.][]{Fritz2018,2018A&A...616A..12G}. Second, to understand the infall and the quenching of star formation in smaller dwarfs, we consider the time of the first infall \citep[e.g.][]{Weisz2015,Fillingham2019}. Third, in order to identify subhaloes brought in by the massive accreted satellite, the radii and velocities of the first and the second most-recent pericentre with respect to the massive accreted satellite are considered \citep[e.g.][]{Patel2020}. These parameters are outlined in Table \ref{table:table1} and demonstrated visually in \ref{fig:fig2}. Throughout this work, we measure and report time in terms of lookback time in Gyr (where the present day is $t=0$\,Gyr). The first pericentre always refers to the most recent pericentre.

\begin{table}
	\caption{A list of the measured parameters of the orbits of the subhaloes. See Fig. \ref{fig:fig2} for a graphic description.}
	\begin{center}
    \begin{tabular}{l|l|l|l|l} 
      \toprule
	  \textbf{With reference to} & \textbf{Position} & \textbf{Rank} & \textbf{Parameter} & \textbf{Code}\\
	  \midrule
	  MW-mass host &  Pericenter & 1 & distance & R\_Per1\\
      & & & time & T\_Per1\\
	  \\
      & Pericenter  & 2 & distance & R\_Per2\\
	  & & & time & T\_Per2\\
	  \\
	  & Pericenter  & 3 & time & T\_Per3\\
	  \\
	  &  Apocenter & 1 & distance & R\_Apo1\\
	  \\
      &  Apocenter & 2 & distance & R\_Apo2\\
	  \\
	  &  Infall & & time & T\_acc\\
      \\
      Massive Progenitor  & Pericenter & 1 & distance & R\_Per1\_Prog\\
	  &  &  & velocity & V\_Per1\_Prog\\
	  \\
      & Pericenter & 2 & distance & R\_Per2\_Prog\\
	  &  &   & velocity  & V\_Per2\_Prog\\
     \bottomrule
	\end{tabular}
	\label{table:table1}
	\end{center}
\end{table}

The distribution of recovered parameters, which allows us to quantify how well we can recover the orbits of the subhaloes, is usually non-Gaussian with extended tails as well as catastrophic outliers. For some purposes, particularly for comparisons where only one headline number is required, we will calculate the root mean square (RMS) deviation  of the recovered parameters. For many other purposes, we wish to present more nuanced parameters, which better reflect the properties of error distributions with large tails towards large deviations. First, we consider that we have failed to recover a parameter $\mathrm{A}$ if: 	
\begin{equation}
\frac{|\mathrm{A_{true}} - \mathrm{A_{est}}|}{\mathrm{A_{true}}} > 0.3.
\end{equation}
that is, if the percentage errors are greater than 30\%, which is our fiducial threshold for accepting a parameter. We calculate an outlier fraction, that is the percentage of subhaloes where we fail to recover a parameter according to the condition above. Second, we express the recovery of the parameters in terms of a calculated mean and scatter (after rejection of outliers). Note that the scatter (as defined) is correlated with the outlier fraction. Furthermore, often the recovered orbit fails to predict a certain pericentre or apocentre (a false negative), while at other times, it also predicts a pericentre or apocentre when none exist in the true orbit of the subhalo (false positive). We calculate the percentage of false positives and false negatives for the detected pericentres/apocentres. These false positive and negative rates give us a lower limit of the failure to reconstruct the orbits. While calculating the RMS, we assign the default value of 0 to false positive and negatives, making the RMS particularly sensitive to the latter. On the other hand, the mean, scatter, outlier rate and the percentage of false positives and negatives allow us to characterise how well we are recovering the orbits for a given model.

\section{Energetics}
\label{sec:energetics}
\begin{figure}
\begin{center}
\includegraphics[trim=0 10 0 0, clip,width=0.9\columnwidth]{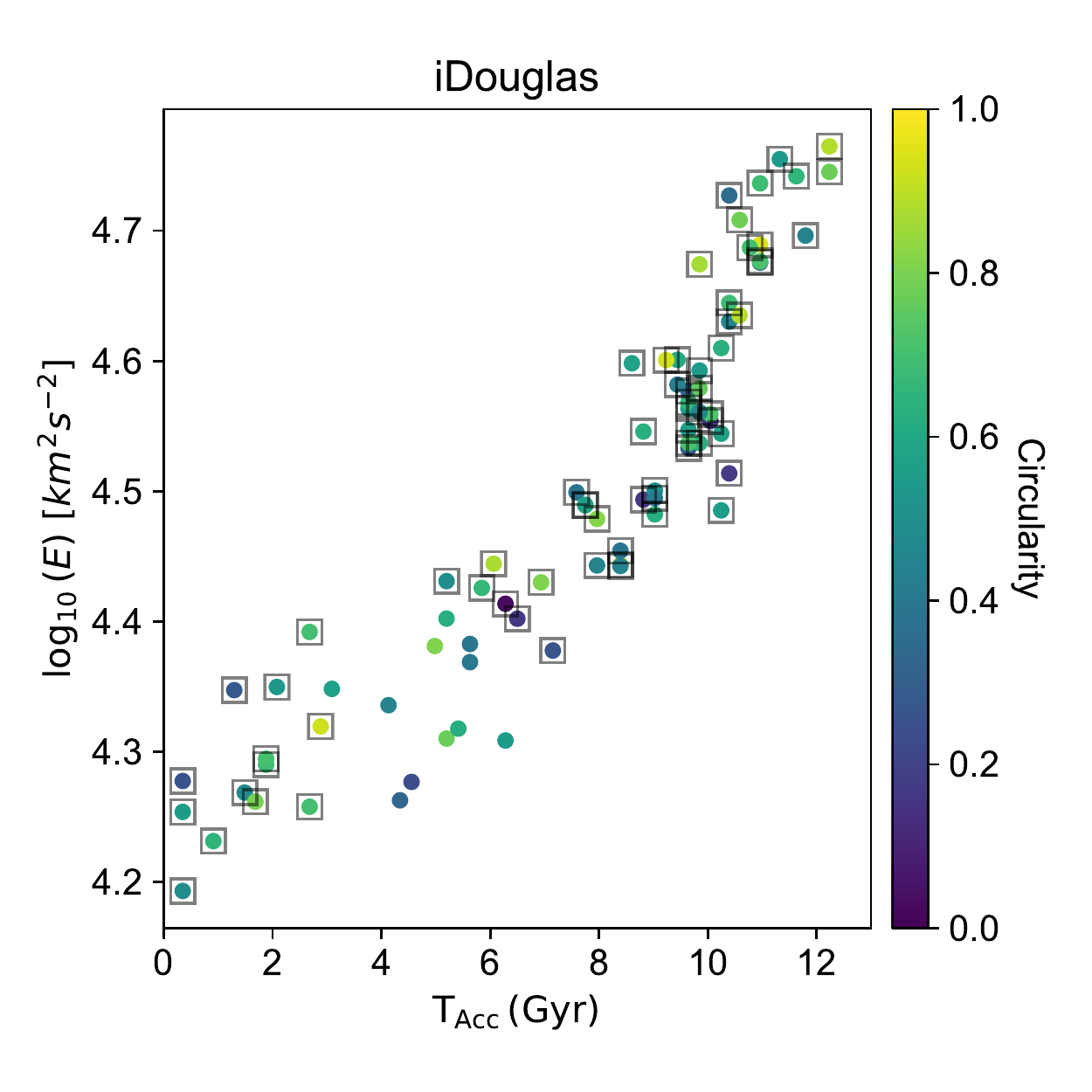}
\includegraphics[trim=0 10 0 0, clip,width=0.9\columnwidth]{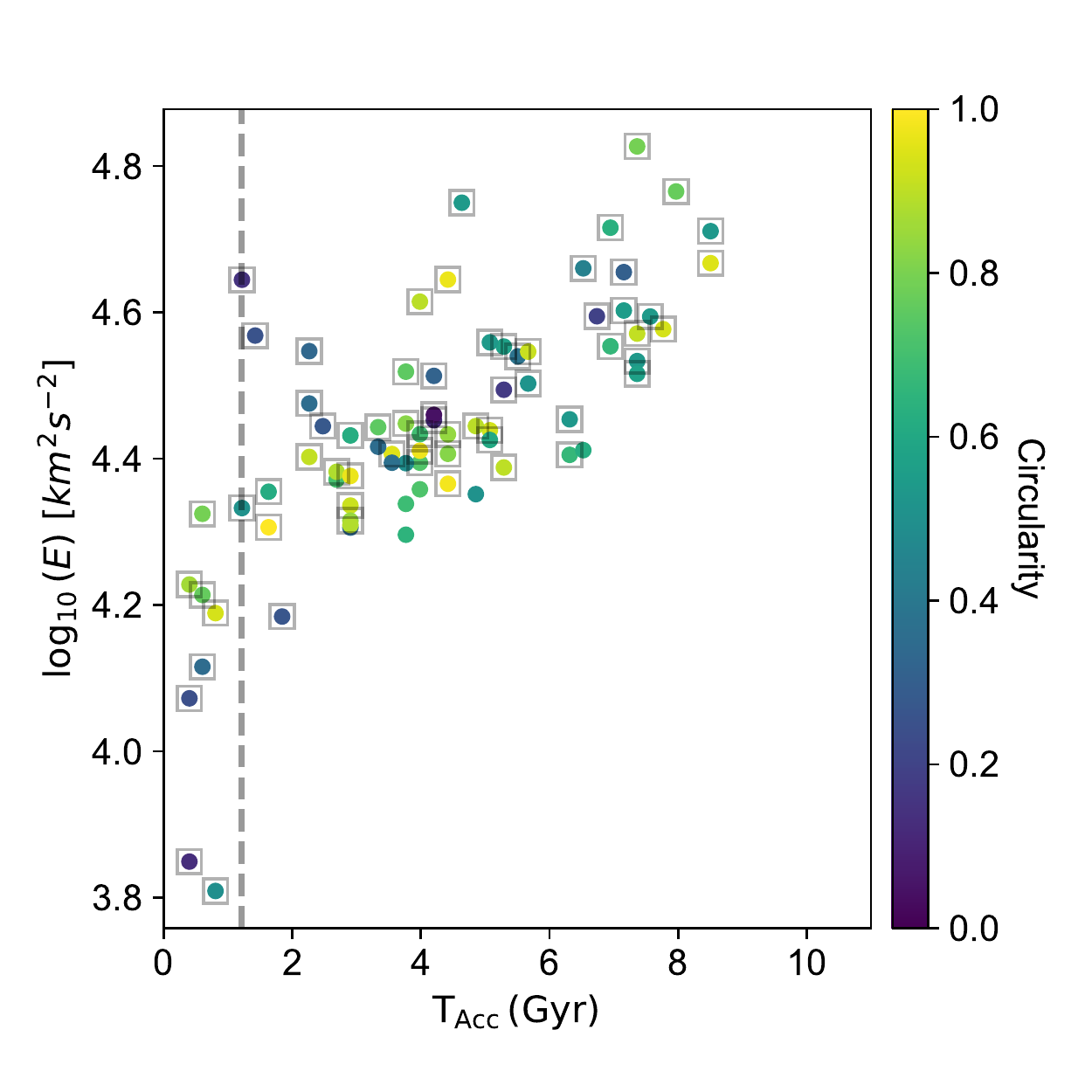}
\caption{The present-day binding energy of the surviving subhaloes at $\mathrm{z=0}$ as a function of the time of accretion for the MW-mass haloes `iDouglas' (top) and `iOates' (bottom). For `iOates', the time of accretion is redefined such that its massive satellite is just past its pericenter. The vertical dashed line represents the accretion time of the massive satellite. The subhaloes are colour-coded by their circularity parameter $\epsilon_J$. Subhaloes marked with a square are presently found within the virial radius. In MW-mass haloes which grow smoothly over time, the present-day binding energy of the subhaloes should be an increasing function of their time of accretion. In the presence of a massive accretion, there is a large scattering in the binding energy. Moreover, the subhaloes accreted along with the massive satellite have a wide range in circularity, or a wide range in angular momentum. Some subhaloes accreted 3-5 Gyr ago are found outside the virial raidus of the main halo.}
\label{fig:fig3}
\end{center}
\end{figure}

Before we attempt to backward integrate the orbits of the subhaloes, we briefly examine the present-day binding energies of the subhaloes of MW-mass haloes `iDouglas' and `iOates' in Fig. \ref{fig:fig3}. For `iDouglas', which did not suffer a massive accretion through out its history, we find that the present-day binding energies of the subhaloes are an increasing function of the time of accretion \citep[See Fig. 3 of][]{Rocha2012}. The scatter in the binding energies is considerably smaller than the overall increase in present-day binding energy with infall time, reflecting the fact the halo grew smoothly across cosmic history. On the other hand, `iOates' recently suffered a 1:10 merger. Although the binding energy of its subhaloes broadly increases with lookback time of accretion, the accretion of subhaloes along with a massive accreted satellite also causes a substantial scatter in their binding energies --- indeed, the scatter is {\it comparable to the full range of binding energies}. `iOates' also accreted similar massive progenitors in the past, in particular one around $\sim$7 Gyr ago, leading to a similar scatter in binding energy at that lookback time. It is clear that with the accretion of massive progenitors or satellites, it is extremely difficult to infer the accretion time of a subhalo from its present-day binding energy. 

\section{Backward integrating Orbits}
\label{sec:backintegrate}
\begin{figure*}
\begin{center}
\includegraphics[trim=80 20 80 20, clip,width=\textwidth]{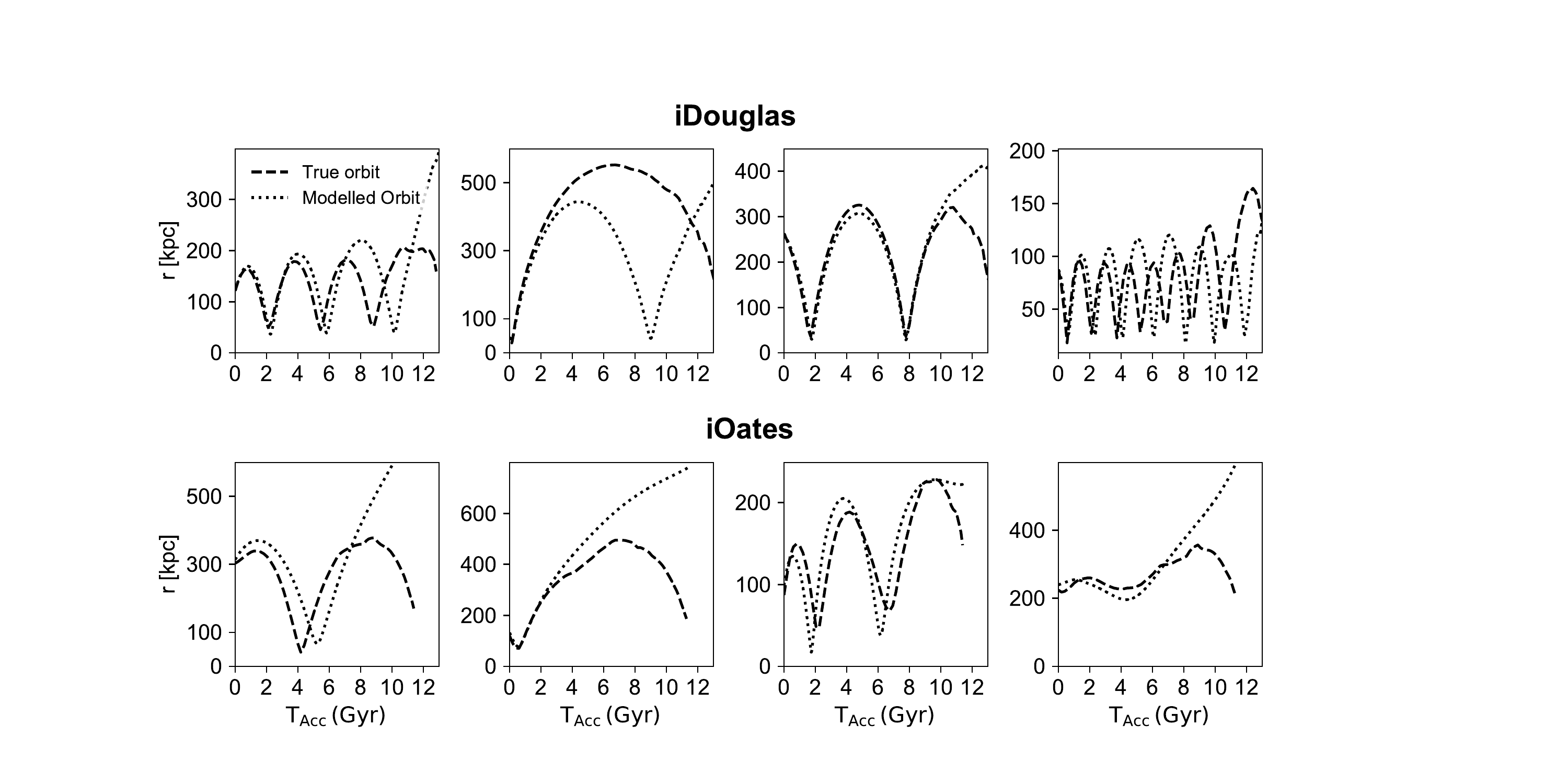}
\caption{Examples of backward integrated orbits of several subhaloes from `iDouglas' and `iOates'. We plot the galactocentric distance (r) as a function of lookback time in Gyr. While `iDouglas' had a quiet accretion history, `iOates' recently accreted a massive satellite, whose potential we also include in the backward integration. The dashed line represents the true orbit from the simulation, while the dotted line represents the backward integrated orbits. In `iDouglas', we reproduce the orbits for the first 3-5 Gyr, after which the backward integrated orbit diverges from the real orbits. In `iOates', differences between the backward integrated orbits and the real orbits appear already quite close to the present day.}
\label{fig:fig4}
\end{center}
\end{figure*}

\begin{figure*}
    \includegraphics[trim=0 90 0 30, clip,width=\textwidth]{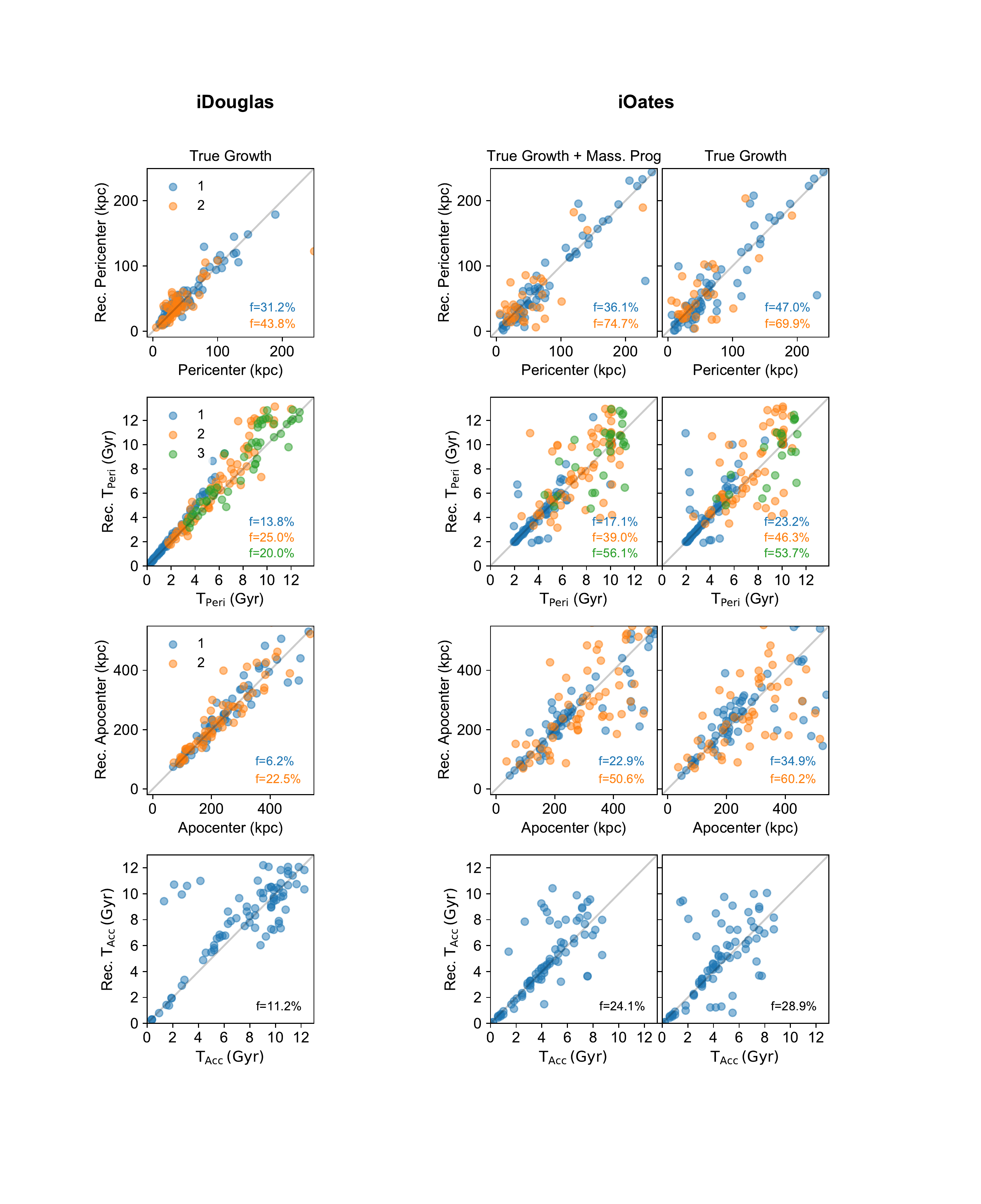}
    \caption{The recovery of various parameters of the orbits of the subhaloes of `iDouglas' (with a quiescent accretion history) and `iOates' (which recently accreted a massive satellite). We integrate assuming the mass growth of the main haloes as reported by the subhalo catalogues. In addition for the case of `iOates' (second column), we integrate by also adding the potential of the recent massive satellite. In the last column, we backward integrate the orbits of the subhaloes of `iOates' using the true mass-growth of the central halo, but neglecting the potential due to the recent massive satellite. The coloured symbols indicate the first (1), second (2) and third (3) pericentre or apocentre. False positives and false negatives are not show in the figure. In the bottom right side of each panel, we indicate the outlier-fraction.}
	\label{fig:fig5}
\end{figure*}

In this section, we attempt to characterise how well we reconstruct the orbital histories of the subhaloes of MW-mass haloes `iDouglas' (without any massive merger) and `iOates' (with a recent massive merger). For `iDouglas', we backward integrate the orbits of the subhaloes using the true mass growth taken from the merger trees. In the case of `iOates' which has recently accreted a massive satellite (~1:10 merger) similar to the accretion of the LMC onto the MW, we backward integrate the orbits of the subhaloes using two different models. In the first model, we include the true mass growth of the central MW-mass halo as well as the potential of the most recent massive accreted satellite, with their respective orbits taken directly from the merger trees. In the second model, we neglect the potential of the most recent massive accreted satellite, while including the true mass growth of the central halo. In order to mimic the particular situation of the MW and the LMC, we start our integration from the snapshot just after the first pericentre of the massive accreted satellite, and calculate lookback time relative to this snapshot. In Fig. \ref{fig:fig4}, we plot some random examples of backward integrated orbits of both `iDouglas' and `iOates', where in the latter we include the potential of the recently accreted massive prognetior. In Fig. \ref{fig:fig5}, we plot the recovery of the orbital parameters vs.\ their true values and display their statistics in Table \ref{table:table2} found in the Appendix \ref{appendix:sec_stat}. We note that false positives and false negatives are not shown in the figure, but are tabulated in Table \ref{table:table2}.

Even in the simple case of `iDouglas' which did not suffer any massive accretion, the backward integrated orbits of individual subhaloes differ slightly from the actual orbits from the simulations. This is clearly demonstrated in Fig. \ref{fig:fig3} where we plot the galactocentric distance of several backward integrated orbits as a function of lookback time. While the recovered orbit agrees with the actual orbit fairly well in the first 4 Gyr, significant differences in the galactocentric distances appear at larger lookback times. This is also reflected in Fig \ref{fig:fig5} where we see the deviations of the recovered parameters from their true values. These deviations are reflected in the single parameter RMS deviations but also in the outlier rates and the false positive/negatives rates found in Table \ref{table:table2} (see Appendix \ref{appendix:sec_stat}). While the scatter in the parameters mirrors their outlier rates, the bias in all the parameters is minimal. If we use the benchmark of the second pericentre, 44\% of the time we fail to predict correctly the pericentric distance to better than 30\%, while 25\% of the time we fail to predict correctly the time of pericentric passage. Moreover, 14\% of the time, the recovered orbits are completely wrong, and fail to predict the second pericentric passage.

In the deviations of the various parameters, we can discern two main effects: First, among similar parameters the errors increase with integration time: e.g., more ancient pericentres are more difficult to recover than more recent pericentres. Second, certain parameters are more affected than others. That is, parameters which depend predominantely on binding energy are easier to recover than parameters that depend on angular momentum. The least biased parameter to recover is the time of accretion, which is a strong function of binding error \citep{Rocha2012}. The pericentric distance (which depends on angular momentum and binding energy) is more difficult to recover than the time of pericentre or the apocentric distance (which depend primarily on binding energy, see Appendix \ref{appendix:sec1}). This suggests that fractional error in the angular momentum is larger than that of the energy in the recovered orbits. That is, while the parametric form of the potential adequately describes the depth of the potential, the deviations in the actual potential from the assumed smooth spherical potential leads to 2nd order random errors in the recovered orbits, which manifest themselves in certain parameters. This suggests that the spherical NFW profile does not provide an adequate description of the real 3D potential of the MW-halo to successfully recover the full orbit of the subhaloes.

The inhomogeneities induced due to the recent accretion of a massive satellite exacerbate the problem. Even if we include the potential of the recent massive satellite for the MW-mass halo `iOates', large differences are seen between the recovered and the actual orbits. While it is difficult to directly compare the recovered parameters of two different MW-mass haloes, the recovered orbits of the subhaloes of `iOates' show significantly larger outlier rates and false positive/negatives among all recovered parameters. Considering again the second pericentre with regards to the MW-mass halo, 75\% of the time we fail to predict correctly the second pericentre distance to better than 30\%, while 40\% of the time we fail to predict correctly the time of the second pericentric passage. Moreover, 23\% of the time, the modelled orbits are completely wrong, and fail to predict the second pericentric passage at all. Other parameters show similar large outlier rates. Part of the scatter found in the temporal parameters (time of accretion and time of pericentric passages) at large lookback times is due to the fact that we have failed to model the previous massive accretions (> 6 Gyr ago). In spite of this, the time of accretion parameter does not exhibit a significant increase in its outlier rates. This suggests that while the overall depth of the potential is well reproduced, our assumed parametric form of the potential, i.e., the sum of the smooth spherical potentials of the MW-mass host and recent massive satellite (as is the state-of-the-art approach for modelling orbits in the Milky Way), fails to accurately recover the full orbit of most of the subhaloes of this system.

The RMS in the recovered parameters is significantly large. Even after removing the false positives/negatives (see Appendix \ref{appendix:sec_stat}), we find considerably large scatter in the recovered parameters. For example, in the case of `iOates', the RMS in the first pericentre is greater than 60 kpc (see Table \ref{table:App_RMS}). Yet from Fig. \ref{fig:fig5}, it appears that the errors are random.  Thus, we conclude that the use of the parametric forms of the potential to backward integrate the orbits of subhaloes leads to significant errors in the recovered parameters.

The failure to model the recent massive accreted satellite leads to much higher outlier and false positive/negatives rates. While there is a general increase in outlier rates among nearly all parameters, the first and second apocentres are particularly affected. This suggests that failure to model the recent massive accreted satellite leads to a misestimation of the depth of the potential during the phase of the accretion of the massive satellite, resulting in large errors in the recovered orbits of the subhaloes \citep{Patel2020,Battaglia2021}.

\subsection{Orbits of subhaloes w.r.t. massive satellite}
\begin{figure*} 
\includegraphics[trim=0 80 0 0, clip,width=0.8\textwidth]{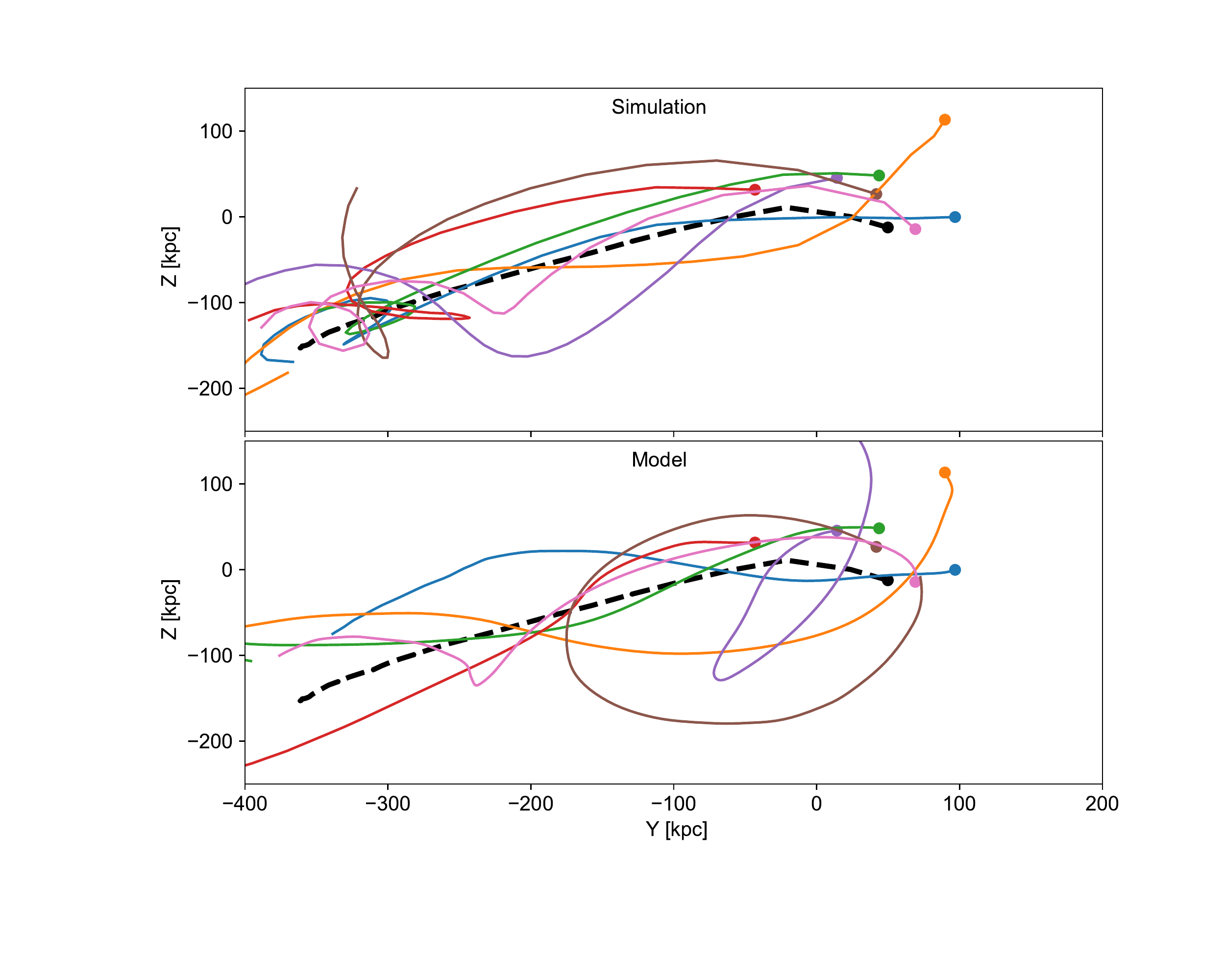}
\caption{Comparison of the backward integrated orbits of some subhaloes of the massive satellite in `iOates' for the past 6 Gyr. The orbits are demonstrated in the reference frame of the MW-mass host; hence the position of the MW-mass host is at (0,0). The black dashed lines represent the orbits of the massive accreted satellite. The top plot represents the orbits from the simulations, while the bottom plot represents the recovered orbits. The backward integrated orbits do not closely resemble the real orbits from the simulations.}
\label{fig:fig6}
\end{figure*}

\begin{figure}
	\begin{center}
	\includegraphics[trim=0 20 0 0, clip,width=0.9\columnwidth]{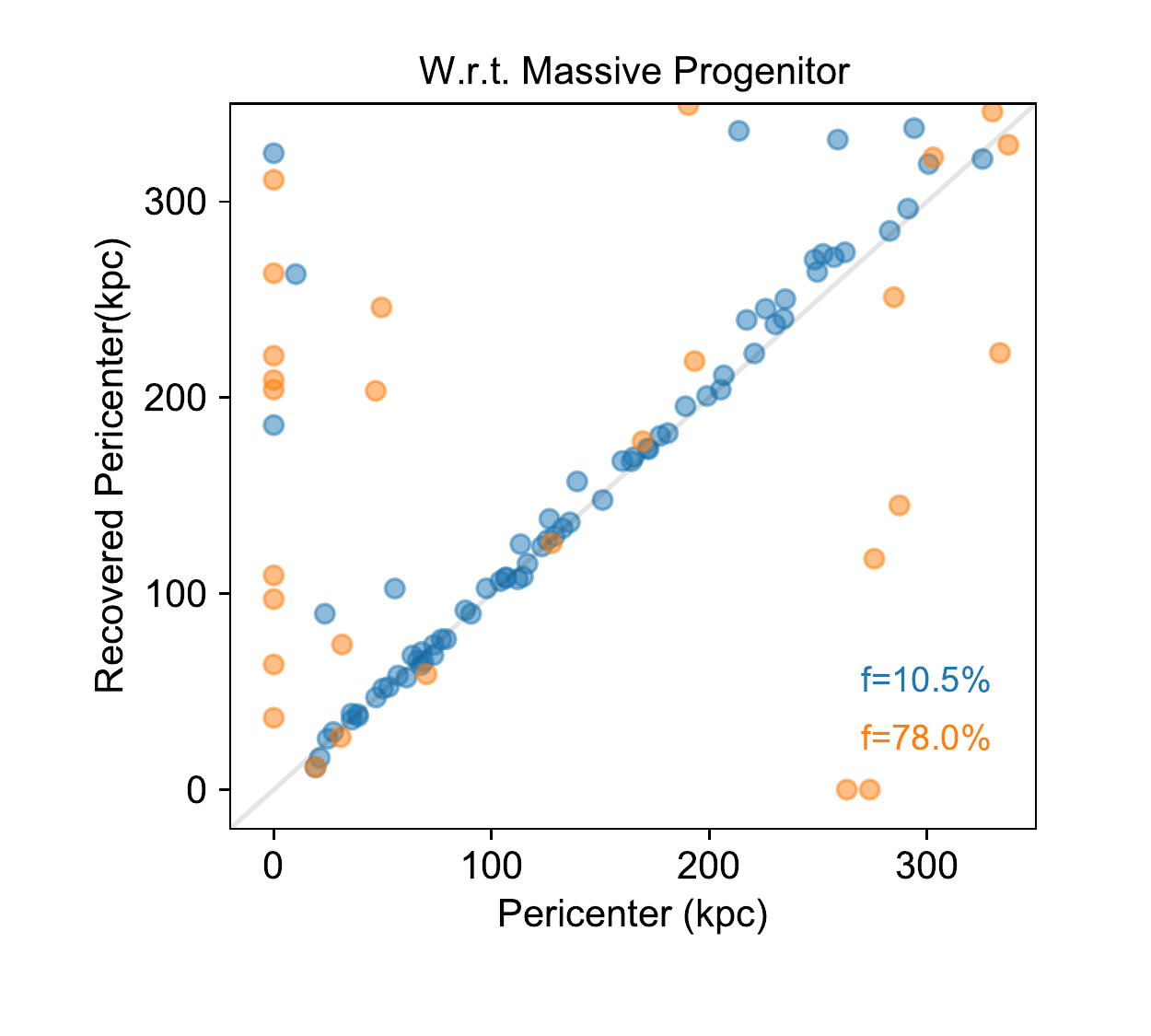}
	\includegraphics[trim=0 20 0 10, clip,width=0.9\columnwidth]{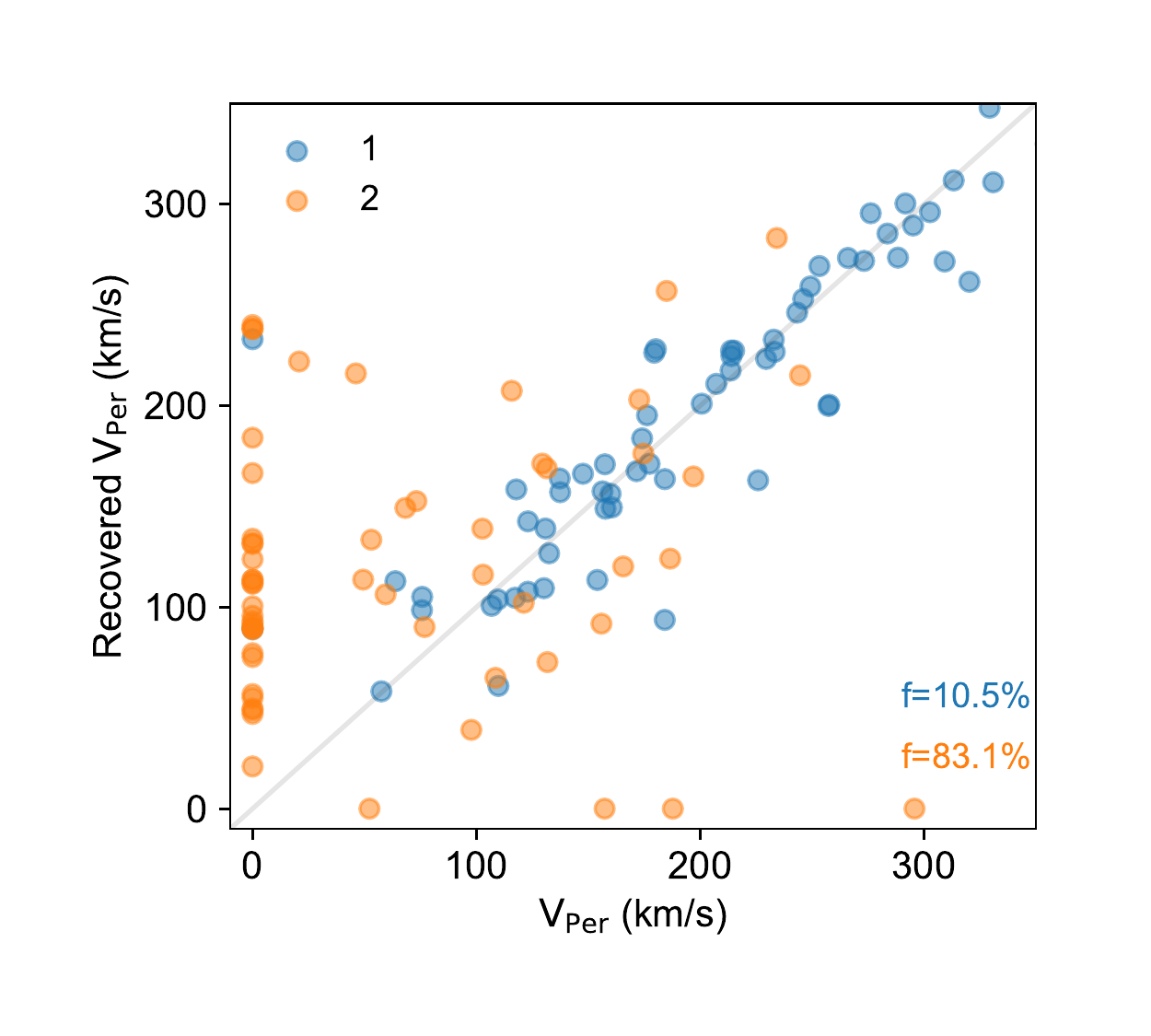}
	\caption{The recovery of the distance and the velocity at the first/second pericentre of the subhalo with respect to the massive satellite. The first and second pericentre are distinguished by different colours and the numbers 1 and 2 respectively. The outlier fraction is shown in the right-lower corner of the figure. While the first pericentre distance and velocity are generally recovered well, the second pericentre is in generally very poorly recovered, and includes many false positives of galaxies that did not have a second pericentre being erroneously assigned one. Note that we explicitly show the false positives (x-axis=0) and the false negatives (y-axis=0) in the figures.}
	\label{fig:fig7}
	\end{center}
\end{figure}

We now turn our attention briefly to understanding the recovery of various parameters of the orbits (distances and velocities at the 1st/2nd pericentre) of subhaloes w.r.t. the massive accreted satellite. These parameters have been used to identify galaxies that might have previously been satellites of the LMC \citep[e.g.][]{Patel2020}. In this subsection, we limit ourselves only to modelling the true mass growth of the central MW-mass host as well as the potential of the massive satellite. In Fig. \ref{fig:fig6}, we compare the recovered orbits of a few subhaloes which were accreted along with the massive satellite along with their true orbits. The recovered orbits appear very different from the true orbits. Furthermore, a fraction of the recovered orbits appear not to be associated with the massive accreted satellite. We quantify this in detail in Fig. \ref{fig:fig7}, where we show how well we recover the distance and velocity of the orbits w.r.t the massive accreted satellite at first/second pericentre. We present the statistics in Table \ref{table:table3} found in Appendix \ref{appendix:sec_stat}. Since we begin our integration of the orbits when the massive satellite is just past pericentre, we can easily recover the distances and velocities at the first pericentre w.r.t the massive satellite. However, our analysis shows that even with complete knowledge of the mass growth of the central MW-mass halo and the exact orbit of the massive satellite, it is {\it still} very difficult to recover the distances and velocities of the second pericentre, with a failure rate greater than 50\% as well as a high percentage of false positives (subhalos without a second pericentre in the simulation that end up with a recovered second pericentre). This suggests that it is very difficult to recover the second pericentre w.r.t. the massive satellite; this may call into question the utility of the practice of using the second pericentre to identify potential satellites of a massive, LMC-like satellite \citep[e.g.][]{Patel2020}.

\section{Comparison of sources of error}
\label{sec:compare}
In this section, we compare the magnitude of the errors in the recovered parameters for the ensemble of the backward integrated orbits of `iOates' due to the use of parametric models of the potential to the other known sources of uncertainty. In particular, we consider four sources of uncertainty considered in the literature: the uncertainty in the present-day virial mass of the central halo, the uncertainty and bias due to not modelling the recent massive accreted satellite, the uncertainty in the initial conditions (the present-day measured 6d phase-space coordinates) and finally, the uncertainty and bias due to not modelling the inertial forces caused by the movement of the halo through the cosmic web. The uncertainty and bias caused by various assumptions of the mass-growth and concentration of the halo are considered in Appendix \ref{appendix:sec4}). These sources of uncertainty may lead to systematic and random errors. Biases in the present-day virial mass of the central halo lead to systematic errors in the recovered parameters. Not modelling the massive satellite leads to both systematic and random errors, where the latter is due to the different phases of the orbits w.r.t. the massive satellite. Uncertainties in the initial conditions leads to random errors in the recovered parameters. We compare the magnitude of the errors using the `outlier fraction', a metric which allows us to gauge the percentage of subhaloes where we fail to accurately recover the parameters to an accuracy better than 30\%. In particular, we calculate the `outlier fraction error', i.e., the outlier fraction between the recovered orbits of the subhaloes compared to the true orbits of the simulation calculated in Section \ref{sec:backintegrate}. We then compare this to `outlier fraction deviation', which compares the parameters obtained by backward integrating the orbits of the orbits of the subhaloes with the given uncertainty and without.

\begin{figure*}
	\includegraphics[trim=0 80 0 30, clip,width=\textwidth,]{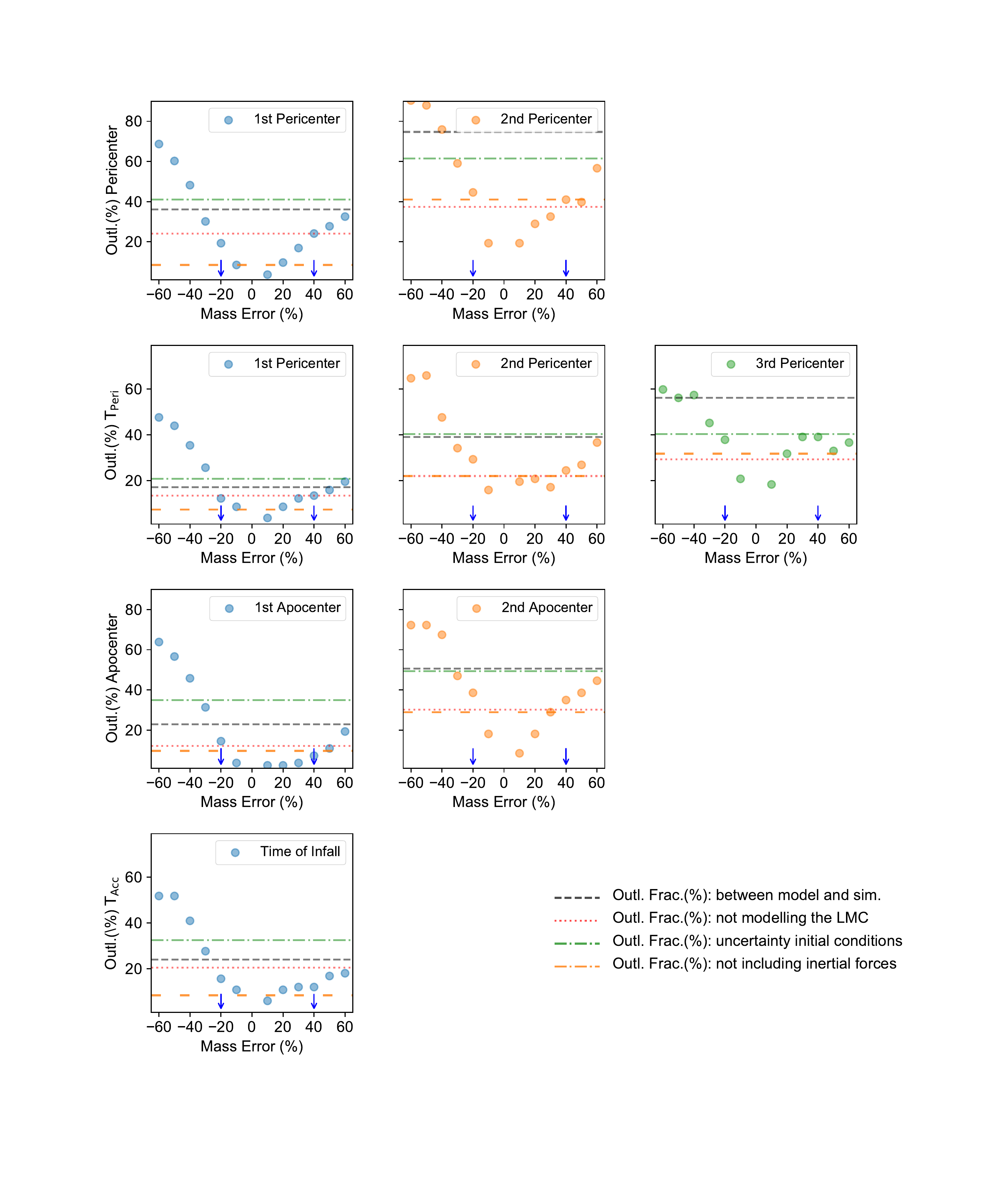}
	\caption{Comparing various sources of error: The outlier fraction deviations of various recovered parameters of `iOates' as a function of changes in the mass of the potential of the central host. The  black dashed horizontal line shows the outlier fraction error between the recovered orbits and the true orbits from the simulation presented in Table \ref{table:table2}. The red dotted horizontal line shows the outlier fraction deviation in the recovered parameters for not modelling the recently accreted massive satellite. The green dot-dashed line shows the outlier fraction deviation due to uncertainties in the initial conditions. We suggest that the outlier fraction deviation due to a -20\% to 40\% uncertainty in the mass of the central MW halo is comparable to the outlier fraction error due to the use of the parametric models of the potential. We indicate using arrows the suggested range.}
	\label{fig:fig8}
\end{figure*}

The uncertainty due to the present-day virial mass of the halo is one of the biggest sources of systematic uncertainties in backward integrating the orbits of satellites. In this section, we consider an overall percentage bias in the virial mass of the halo at all redshifts, while following the true growth history of the halo. The first thing we notice is that the outlier fraction deviations in the recovered parameters are asymmetric around positive and negative changes in the mass of the central potential, i.e, the outlier fraction deviations in the recovered parameters due to a decrease in the mass of the potential are much higher than those due to a corresponding increase in the mass of the potential. This is because larger the mass of the potential, the longer a subhalo orbits around the central MW-mass host. However, smaller the mass of the potential, the more easier it is for subhaloes to escape the gravitational field of the central host - resulting in a larger number of false negatives and a corresponding higher cost for the outlier fraction deviation metric. This results in asymmetric outlier fraction deviations of the recovered parameters for changes in the mass of the potential, with smaller masses penalised more than larger masses. We find that the outlier fraction errors of the parameters of the recovered orbits (due to the use of parametric models for the potential) are comparable to outlier fraction deviations caused by a 20\% decrease or a $\sim40-50$\% increase in the mass of the potential, which is equivalent to a $\pm$30\% change in the mass of the potential. The first and second pericentre are the only parameters to show higher outlier fractions errors compared to the other parameters - owing to their very small absolute values and the definition of the outlier fraction. Moreover, Fig. \ref{fig:fig8} reminds us that we can recover the parameters to better than 30\% for at most 60 to 80\% of the subhaloes. We conclude that in general, the magnitude of the errors in the recovered parameters due to the use of the parametric forms of the potential is comparable to the errors caused by a 30\% uncertainty in the virial mass of the MW.

Not modelling the potential of the most recent massive satellite is another source of systematic uncertainty. We find that the total outlier fraction error due to the use of parametric models of the potential is comparable or slightly smaller (10-45\% less) to the outlier fraction deviations for not modelling the recently accreted LMC-like massive satellite. Due to the nature of the outlier fraction deviation curve, the uncertainty in the recovered parameters due to not modelling the recently-accreted massive satellite is roughly comparable to outlier fraction deviations caused by a 20\% decrease or a $\sim40-50$\% increase in the mass of the potential. We conclude that the `penalty' of not modelling the recently accreted massive satellite is comparable to the cost of using parametric forms of the potential.

An important source of statistical errors in the recovered parameters of the orbits of MW satellites are the uncertainties in their initial conditions (proper motions and distances). In general, the more massive the MW satellite, the smaller the errors in the proper motions, while distance errors dominate in MW satellites found at larger galactocentric distances \citep[e.g.][]{Battaglia2021}. In order to get a sense of the relative magnitude of these systematic errors due to the uncertainties in the initial conditions of the MW satellites in contrast to the errors due to the uncertainties in the potential, we consider typical errors found in the initial conditions of MW satellites \citep{Battaglia2021}: a 2\% error in each component of the proper motion and a 5\% error in the distances, while neglecting differences in the masses of the various subhaloes. We find that the outlier fraction deviations due to uncertainties only in the initial conditions are comparable to the errors due to the use of the parametric form of the potential. However, we expect that the uncertainties in the initial conditions will decrease with newer data.

Another source of uncertainty is the inertial forces due to the movement of the halo through the cosmic web. The MW is not stationary but is moving due to the forces exerted by large scale structure of the Universe. In fact, one can differentiate between two separate movements: the reflex motion of the MW due to the accretion of the LMC within the last $\sim$1 Gyr, and the overall movement of the MW through the cosmic web \citep[e.g.][]{Nusser2011,Hoffman2015,Hoffman2017} for the last $\sim$13 Gyr. Constaints on the latter contain a high degree of uncertainty and are usually not included in backward integration of MW satellites. In this work, we can separate the two motions because of the way in which we choose to model the MW-mass halo and the recently accreted massive satellite. That is, we first model the exact motion of the MW-mass halo and the recently accreted massive satellite taken directly from the ROCKSTAR catalogues as moving potentials. Secondly, we consider a moving frame of reference close to the MW-mass halo by fitting a 4th order spline to its motion. The inertial forces due to the motion of the moving frame of reference parallels the inertial forces due to the movement of the MW through the cosmic web. The neglect of the inertial forces due to the overall movement of the MW-mass halo through the cosmic web leads to significant errors in the recovered orbits. Not including these inertial forces leads to outlier fraction deviations comparable to the uncertainty in the initial conditions especially for parameters involving the second and third pericentre/apocentre.

Apart from the sources of uncertainty considered here, the triaxility of the halo will also play a major role. In this work we only consider halos with spherical symmetry. The ROCKSTAR catalogues with which form the basis of this work do not report whether the halo is prolate or oblate. Hence we have not direct access to the triaxility of the central halo. Understanding the source of uncertainty due to triaxiality is beyond the scope of this paper.

In this section, we compared various sources of error using the metric of the outlier fraction. However, we must remember that the comparison of the various sources of error is conditioned by the choice of metric we use. While the outlier fraction deals well with false positives/negatives, it fails to take into account the scatter of the parameters with deviations less than 30\%. To test that our results are independent of the choice of the metric used, we repeat our comparisons in Appendix \ref{appendix:sec4} using the RMS metric and obtain similar results. We therefore conclude that the errors caused by the contribution to the total error budget due the use of the parametric forms of the potential rival i) a 30\% uncertainty in the virial mass of the MW as well as ii) not modelling the potential of the recently  accreted massive satellite.

\section{Summary and Discussion}
\label{sec:conc}
This work demonstrates that an important source of error previously unconsidered in reconstructing the orbits of satellites is the use of simple parametric models, which often fail to capture the inhomogeneities present in the MW's halo. We demonstrate this by considering two MW-mass haloes taken from DM-only simulations: a) an isolated MW-mass halo which did not suffer any massive accretion during its history and b) a MW-mass halo which has recently accreted a massive satellite similar to the case of MW and the LMC. We approximate the potential of the central MW-mass halo and the recently accreted massive satellite with a spherical NFW profile and Hernquist profile respectively. We further characterised the recovered orbits through various parameters, and compare them with the true parameters from the simulations. We find that even in the case of an isolated MW-mass halo, small deviations in the potential from the parametric form result in a significant scatter and outlier fraction in the recovered parameters of the orbits of the subhaloes. In the presence of a recently accreted massive satellite, the spatial and temporal inhomogeneities in the potential become even more pronounced and have a much larger effect on the recovered orbits - resulting in large scatter, outlier fraction and false positives/negatives in the recovered parameters. This suggests that simple symmetric parametric forms of the potential, i.e., modelling a recent massive accretion like that of the LMC as a sum of two spherical parametric potentials, fail to capture the complexities and the inhomogeneities of the true potential experienced by the subhaloes and lead to large errors in the recovered parameters of the orbits. Furthermore, we find that the magnitude of the errors are as large as a) the errors caused by a 30\% uncertainty in the virial mass of the central halo and b) the errors caused by not modelling the recently accreted massive satellite.

While convenient, simple parametric forms of the potential fail to capture the small-scale deviations and inhomogeneities of the potential as well as its global shape. Simple parametric models help define the depth of the potential with only a few parameters and constraints. On the other hand, they neglect any clumpiness, triaxility, twistedness and possible rotation of the halo. It has been recently demonstrated that high fidelity orbit reconstruction can be obtained by modelling the changing complex potential of a simulated isolated MW-mass galaxy using a multipole expansion at $\sim$100 Myr temporal resolution \citep{Sanders2020}. The recent passage of a massive accreted satellite complicates the situation further due to the resonances induced in the central DM halo: large transient density and kinematic perturbations \citep{Choi2009} are produced, creating the classical `conic' wake trailing the satellite \citep{Chandrasekhar1943} as well as a collective response to the system as a whole \citep{Garavito-Camargo2019} and leading to a temporary change in the concentration of the halo \citep{Wang2020}. In fact, it has been recently demonstrated that the LMC induces a reflex motion in the disk/halo of the MW \citep{Gomez2015,Erkal2021,Petersen2021} and a large wake in the halo of the MW \citep{Conroy2021}. While modelling the wake and the other homogeneities is beyond the scope of this paper, our work suggests that the errors caused by the use of these spherical parametric models for the potential rival the 30\% uncertainty in the virial mass of the MW, and is thus an important source of error to consider. Recently, \cite{Garavito-Camargo2021} characterised the inhomogeneities induced in the MW halo due to the infall of the LMC using basis functions, while \cite{Cunningham2020} has demonstrated that a spherical harmonic expansion of the stellar halo velocity field may allow us to quantify the response of the MW DM halo to the LMC's infall. Upcoming spectroscopic campaigns of halo stars may allow us to model with greater fidelity the present-day inhomogeneities in the potential of the MW.

Until such models are widely available, failure to take into account the errors due to the use of parametric forms of the potential may lead to under-estimated uncertainties in some of the recovered parameters of the orbits of the MW satellites. In addition to the three primary sources of error previously considered in the literature (the uncertainty in the mass of the central MW-mass halo, the uncertainty due to not modelling the recently accreted massive satellite and the uncertainty in the initial conditions), one has to additionally consider the uncertainty in the recovered parameters due to the use of the parametric models of the potential. While some uncertainties are systematic, other are random, and they combine in non-linear ways. Furthermore, as our constraints on the initial conditions and the mass of the MW become better with time, the errors due to our use of parametric forms of the potential will be dominant source of error. This work suggests that the uncertainties in some of the recovered parameters of the orbits of several MW satellites may be under-estimated, if one fails to account for the errors due to the use of the parametric models.

Furthermore, a primary assumption used in this work is that it is enough to model only the potential of the central MW-mass host and the most recently accreted massive satellite. Our results confirm that it is important to model the potential of the most recently accreted massive satellite. Failure to do so can can produce errors in the recovered parameters of the orbits of the satellites of the MW \citep[e.g.][]{Patel2020}, which can be as large as the errors caused by using the parametric forms of the potential. However, in restricting the potential only to these two large bodies, we neglect the interactions of subhaloes among themselves or with a previously accreted massive progenitor. Yet, in CDM, subhaloes are often accreted in groups. Furthermore, it is assumed that in the case of the MW, a number of satellites may have been accreted with its previous accretions, i.e., the Sagittarius and Gaia-Enceladus galaxies. Yet, it is often difficult to include the potential of previously accreted massive progenitors. First, the exact orbits of these previous massive progenitors is highly uncertain. Second, the errors in the recovered orbit of the subhaloes increase with integration time, making it difficult to model the reflex motion as well the exact 3D interactions with the previously accreted massive progenitors and the central MW-mass host. While it is vitally important to model the LMC, for some applications it may be equally important to model previous massive accretions to be able to successfully backward integrate the orbits of the MW satellites.

As with the other sources of error, the use of parametric forms of the potential affect the recovery of the parameters of the orbits in two complementary ways. First, the longer the integration time of the orbit, the greater the error in the recovered parameter. The errors in the recovered orbit of the subhaloes accumulate with integration time, causing a dramatic effect on parameters measured at large lookback times. For example, the second pericentric distance is much more difficult to recover than the first pericentric distance. Such accumulated errors in the recovered orbit make it is very difficult to model the reflex motion of previous massive accretions. Second, different `types' of parameters have different sources of error. In general, the errors in a particular type of parameter correlate with the number of phase-space-time coordinates involved in the measurement of that type of parameter. For example, time of pericentric passage is much easier to measure than the actual pericentric distance. Another example of this is that apocentres (which depend primarily on binding energy) are easier to recover than pericentres (which depend on both binding energy and angular momentum). Both principles affect the recovery of the individual parameters of the orbits, and their consequences are not often very obvious. We highlight two surprising yet important consequences. First, it is very difficult to recover the distance and the velocity at the second pericentre w.r.t the recently accreted massive satellite, a parameter important to determine which satellites came in with the LMC. Second, in the absence of systematic biases (e.g. incorrect virial mass of the central MW-mass halo or not modelling the recently accreted massive satellite), the time of accretion of the subhalo is one of the easiest parameters to recover, and reflects the fact that in the absence of massive accretions, the time of accretion is a function of binding energy \citep{Rocha2012}.

\begin{figure}
\includegraphics[trim=0 10 0 0, clip,width=\columnwidth]{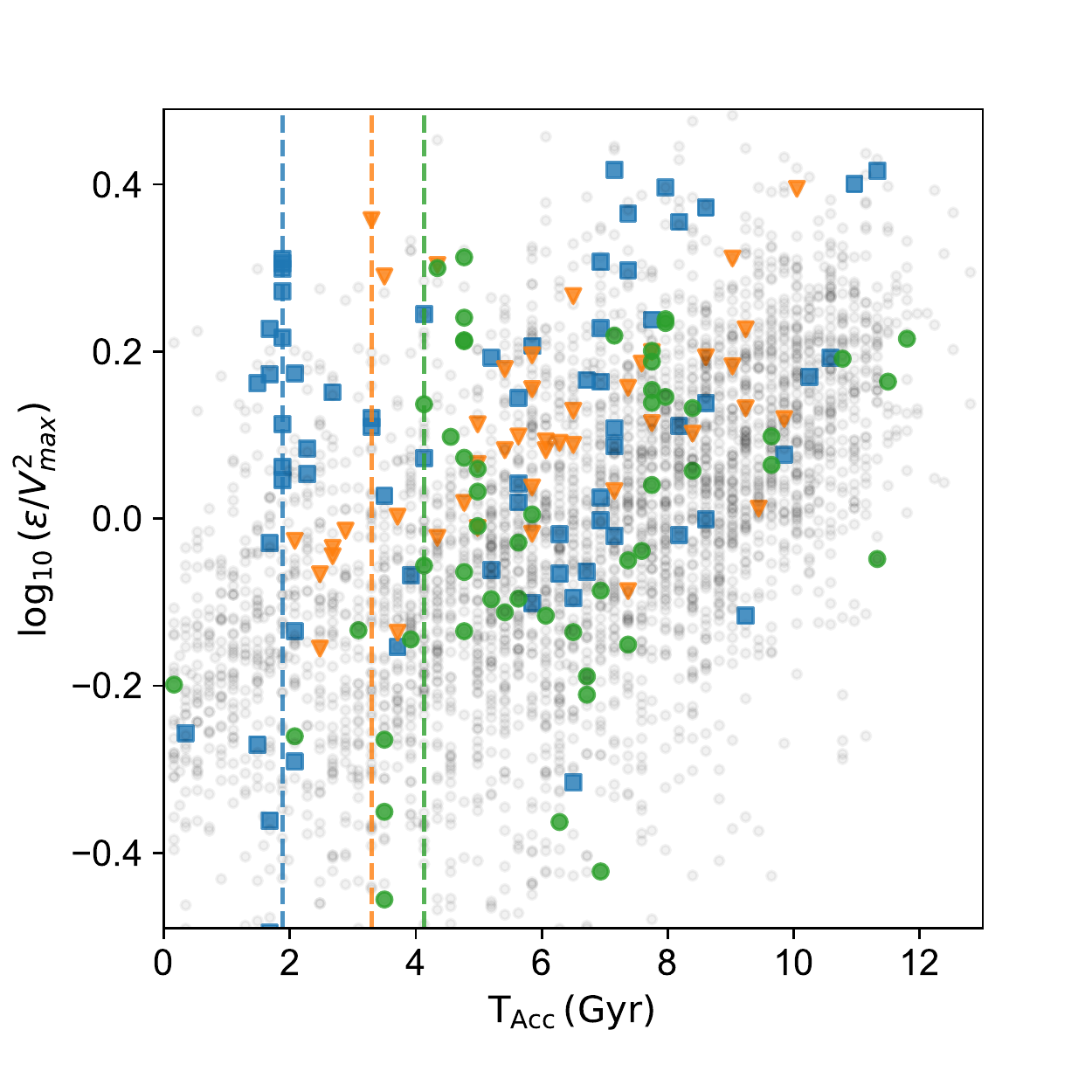}
\caption{The present-day binding energies of the subhaloes of all 48 MW-mass haloes in the ELVIS simulation suite as a function of their infall time. We highlight in different colours the three particular MW-mass haloes which have recently accreted massive satellites, and which closely resemble the recent accretion of an LMC. In general, the present-day binding energy of the subhaloes increases as a function of their infall time. However, in the case of a large massive accretion, there is a scattering of the binding energy of those subhaloes which are accreted at the same time as the large massive accretion.}
\label{fig:fig9}
\end{figure}

Yet, in the presence of a massive sattellite, it is difficult to infer the infall time of MW satellites from their present-day binding energies as proposed by \cite{Rocha2012} and \cite{Fillingham2019}. The accretion of a massive satellite or progenitor causes a large scattering in the binding energies of the subhaloes accreted contemporaneously, with the amplitude of the scatter often rivalling the total range in the binding energy of the subhaloes. With the accretion of a massive satellite or progenitor, the present-day binding energies of subhaloes is no longer a steadily increasing function of infall time. In Fig. \ref{fig:fig9}, we plot the present-day binding energies of the subhaloes of all 48 MW-mass haloes included in the ELVIS suite \citep{Garrison2014} as a function of their infall times. In general, the present-day binding energy of the subhaloes increases (with substantial scatter) as a function of their lookback accretion time (as broadly expected by \citealt{Rocha2012}). However, highlighting the three particular MW-mass hosts in the ELVIS suite which have suffered recent LMC-like accretions (Fig. \ref{fig:fig9}), we can see that these accretions lead to a much higher scatter in their present-day binding energies. The scattering in binding energy appears most prominently just after the massive satellite/progenitor has reached pericentre. Importantly, this implies that the infall time of MW satellites cannot be directly inferred from their present-day binding energies, due to the ongoing accretion of the LMC.

Finally, the recovery of the parameters of the orbits of MW dwarf galaxies will also depend upon the variety of assumptions that goes into the backward integration. Furthermore, the uncertainties in the initial conditions of individual MW satellites will depend on its mass and its distance. For this reason, it is difficult to comment on results of backward integrations of individual MW satellites, and the associated error with the use of parametric forms of the potential. This work suggests that it is important that researchers test their backward integrations on cosmological simulations which include a recently accreted massive satellite in order to characterise the  uncertainty associated with their specific techniques and assumptions of the potential.

\section*{Acknowledgements}
We are grateful to Shea Garrison-Kimmel for providing us access to the catalogues and merger trees of the ELVIS simulations. E.F.B. is grateful for support from the National Science Foundation through grant NSF-AST 2007065 and NASA grant NNG16PJ28C through subcontract from the University of Washington as part of the \textit{WFIRST} Infrared Nearby Galaxies Survey.

%%%%%%%%%%%%%%%%%%%%%%%%%%%%%%%%%%%%%%%%%%%%%%%%%%
\section*{Data Availability}
The catalogues and the merger trees of the simulations analysed in this article were provided by Shea Garrison-Kimmel. These will be shared on request to the corresponding author with permission of Shea Garrison-Kimmel.



\bibliographystyle{mnras}
\bibliography{energetics} % if your bibtex file is called example.bib

\begin{thebibliography}{}
\makeatletter
\relax
\def\mn@urlcharsother{\let\do\@makeother \do\$\do\&\do\#\do\^\do\_\do\%\do\~}
\def\mn@doi{\begingroup\mn@urlcharsother \@ifnextchar [ {\mn@doi@}
  {\mn@doi@[]}}
\def\mn@doi@[#1]#2{\def\@tempa{#1}\ifx\@tempa\@empty \href
  {http://dx.doi.org/#2} {doi:#2}\else \href {http://dx.doi.org/#2} {#1}\fi
  \endgroup}
\def\mn@eprint#1#2{\mn@eprint@#1:#2::\@nil}
\def\mn@eprint@arXiv#1{\href {http://arxiv.org/abs/#1} {{\tt arXiv:#1}}}
\def\mn@eprint@dblp#1{\href {http://dblp.uni-trier.de/rec/bibtex/#1.xml}
  {dblp:#1}}
\def\mn@eprint@#1:#2:#3:#4\@nil{\def\@tempa {#1}\def\@tempb {#2}\def\@tempc
  {#3}\ifx \@tempc \@empty \let \@tempc \@tempb \let \@tempb \@tempa \fi \ifx
  \@tempb \@empty \def\@tempb {arXiv}\fi \@ifundefined
  {mn@eprint@\@tempb}{\@tempb:\@tempc}{\expandafter \expandafter \csname
  mn@eprint@\@tempb\endcsname \expandafter{\@tempc}}}

\bibitem[\protect\citeauthoryear{{Abadi}, {Navarro}, {Steinmetz}  \&
  {Eke}}{{Abadi} et~al.}{2003}]{Abadi2003}
{Abadi} M.~G.,  {Navarro} J.~F.,  {Steinmetz} M.,   {Eke} V.~R.,  2003, \mn@doi
  [\apj] {10.1086/378316}, \href
  {https://ui.adsabs.harvard.edu/abs/2003ApJ...597...21A} {597, 21}

\bibitem[\protect\citeauthoryear{{Battaglia}, {Taibi}, {Thomas}  \&
  {Fritz}}{{Battaglia} et~al.}{2021}]{Battaglia2021}
{Battaglia} G.,  {Taibi} S.,  {Thomas} G.~F.,   {Fritz} T.~K.,  2021, arXiv
  e-prints, \href {https://ui.adsabs.harvard.edu/abs/2021arXiv210608819B} {p.
  arXiv:2106.08819}

\bibitem[\protect\citeauthoryear{{Behroozi}, {Wechsler}  \& {Wu}}{{Behroozi}
  et~al.}{2013a}]{Rockstar}
{Behroozi} P.~S.,  {Wechsler} R.~H.,   {Wu} H.-Y.,  2013a, \mn@doi [\apj]
  {10.1088/0004-637X/762/2/109}, \href
  {https://ui.adsabs.harvard.edu/abs/2013ApJ...762..109B} {762, 109}

\bibitem[\protect\citeauthoryear{{Behroozi}, {Wechsler}, {Wu}, {Busha},
  {Klypin}  \& {Primack}}{{Behroozi} et~al.}{2013b}]{Consistent_trees}
{Behroozi} P.~S.,  {Wechsler} R.~H.,  {Wu} H.-Y.,  {Busha} M.~T.,  {Klypin}
  A.~A.,   {Primack} J.~R.,  2013b, \mn@doi [\apj]
  {10.1088/0004-637X/763/1/18}, \href
  {https://ui.adsabs.harvard.edu/abs/2013ApJ...763...18B} {763, 18}

\bibitem[\protect\citeauthoryear{{Belokurov}, {Erkal}, {Evans}, {Koposov}  \&
  {Deason}}{{Belokurov} et~al.}{2018}]{Belokurov2018}
{Belokurov} V.,  {Erkal} D.,  {Evans} N.~W.,  {Koposov} S.~E.,   {Deason}
  A.~J.,  2018, \mn@doi [\mnras] {10.1093/mnras/sty982}, \href
  {https://ui.adsabs.harvard.edu/abs/2018MNRAS.478..611B} {478, 611}

\bibitem[\protect\citeauthoryear{{Besla}, {Kallivayalil}, {Hernquist},
  {Robertson}, {Cox}, {van der Marel}  \& {Alcock}}{{Besla}
  et~al.}{2007}]{Besla2007}
{Besla} G.,  {Kallivayalil} N.,  {Hernquist} L.,  {Robertson} B.,  {Cox} T.~J.,
   {van der Marel} R.~P.,   {Alcock} C.,  2007, \mn@doi [\apj]
  {10.1086/521385}, \href
  {https://ui.adsabs.harvard.edu/abs/2007ApJ...668..949B} {668, 949}

\bibitem[\protect\citeauthoryear{{Boldrini} \& {Bovy}}{{Boldrini} \&
  {Bovy}}{2021}]{Boldrini2021}
{Boldrini} P.,  {Bovy} J.,  2021, arXiv e-prints, \href
  {https://ui.adsabs.harvard.edu/abs/2021arXiv210609419B} {p. arXiv:2106.09419}

\bibitem[\protect\citeauthoryear{{Bovy}}{{Bovy}}{2015}]{Bovy2015}
{Bovy} J.,  2015, \mn@doi [\apjs] {10.1088/0067-0049/216/2/29}, \href
  {https://ui.adsabs.harvard.edu/abs/2015ApJS..216...29B} {216, 29}

\bibitem[\protect\citeauthoryear{{Bovy}, {Bahmanyar}, {Fritz}  \&
  {Kallivayalil}}{{Bovy} et~al.}{2016}]{Bovy2016}
{Bovy} J.,  {Bahmanyar} A.,  {Fritz} T.~K.,   {Kallivayalil} N.,  2016, \mn@doi
  [\apj] {10.3847/1538-4357/833/1/31}, \href
  {https://ui.adsabs.harvard.edu/abs/2016ApJ...833...31B} {833, 31}

\bibitem[\protect\citeauthoryear{{Bryan} \& {Norman}}{{Bryan} \&
  {Norman}}{1998}]{Byran1998}
{Bryan} G.~L.,  {Norman} M.~L.,  1998, \mn@doi [\apj] {10.1086/305262}, \href
  {https://ui.adsabs.harvard.edu/abs/1998ApJ...495...80B} {495, 80}

\bibitem[\protect\citeauthoryear{{Bryan}, {Kay}, {Duffy}, {Schaye}, {Dalla
  Vecchia}  \& {Booth}}{{Bryan} et~al.}{2013}]{Bryan2013}
{Bryan} S.~E.,  {Kay} S.~T.,  {Duffy} A.~R.,  {Schaye} J.,  {Dalla Vecchia} C.,
    {Booth} C.~M.,  2013, \mn@doi [\mnras] {10.1093/mnras/sts587}, \href
  {https://ui.adsabs.harvard.edu/abs/2013MNRAS.429.3316B} {429, 3316}

\bibitem[\protect\citeauthoryear{{Buist} \& {Helmi}}{{Buist} \&
  {Helmi}}{2015}]{Buist2015}
{Buist} H. J.~T.,  {Helmi} A.,  2015, \mn@doi [\aap]
  {10.1051/0004-6361/201526203}, \href
  {https://ui.adsabs.harvard.edu/abs/2015A&A...584A.120B} {584, A120}

\bibitem[\protect\citeauthoryear{{Bullock}, {Kolatt}, {Sigad}, {Somerville},
  {Kravtsov}, {Klypin}, {Primack}  \& {Dekel}}{{Bullock}
  et~al.}{2001}]{Bullock2001}
{Bullock} J.~S.,  {Kolatt} T.~S.,  {Sigad} Y.,  {Somerville} R.~S.,  {Kravtsov}
  A.~V.,  {Klypin} A.~A.,  {Primack} J.~R.,   {Dekel} A.,  2001, \mn@doi
  [\mnras] {10.1046/j.1365-8711.2001.04068.x}, \href
  {https://ui.adsabs.harvard.edu/abs/2001MNRAS.321..559B} {321, 559}

\bibitem[\protect\citeauthoryear{{Callingham} et~al.,}{{Callingham}
  et~al.}{2019}]{Callingham2019}
{Callingham} T.~M.,  et~al., 2019, \mn@doi [\mnras] {10.1093/mnras/stz365},
  \href {https://ui.adsabs.harvard.edu/abs/2019MNRAS.484.5453C} {484, 5453}

\bibitem[\protect\citeauthoryear{{Cautun} et~al.,}{{Cautun}
  et~al.}{2020}]{Cautun2020}
{Cautun} M.,  et~al., 2020, \mn@doi [\mnras] {10.1093/mnras/staa1017}, \href
  {https://ui.adsabs.harvard.edu/abs/2020MNRAS.494.4291C} {494, 4291}

\bibitem[\protect\citeauthoryear{{Chandrasekhar}}{{Chandrasekhar}}{1943}]{Chandrasekhar1943}
{Chandrasekhar} S.,  1943, \mn@doi [\apj] {10.1086/144517}, \href
  {https://ui.adsabs.harvard.edu/abs/1943ApJ....97..255C} {97, 255}

\bibitem[\protect\citeauthoryear{{Choi}, {Weinberg}  \& {Katz}}{{Choi}
  et~al.}{2009}]{Choi2009}
{Choi} J.-H.,  {Weinberg} M.~D.,   {Katz} N.,  2009, \mn@doi [\mnras]
  {10.1111/j.1365-2966.2009.15556.x}, \href
  {https://ui.adsabs.harvard.edu/abs/2009MNRAS.400.1247C} {400, 1247}

\bibitem[\protect\citeauthoryear{{Conroy}, {Naidu}, {Garavito-Camargo},
  {Besla}, {Zaritsky}, {Bonaca}  \& {Johnson}}{{Conroy}
  et~al.}{2021}]{Conroy2021}
{Conroy} C.,  {Naidu} R.~P.,  {Garavito-Camargo} N.,  {Besla} G.,  {Zaritsky}
  D.,  {Bonaca} A.,   {Johnson} B.~D.,  2021, arXiv e-prints, \href
  {https://ui.adsabs.harvard.edu/abs/2021arXiv210409515C} {p. arXiv:2104.09515}

\bibitem[\protect\citeauthoryear{{Contigiani}, {Rossi}  \&
  {Marchetti}}{{Contigiani} et~al.}{2019}]{Contigiani2019}
{Contigiani} O.,  {Rossi} E.~M.,   {Marchetti} T.,  2019, \mn@doi [\mnras]
  {10.1093/mnras/stz1547}, \href
  {https://ui.adsabs.harvard.edu/abs/2019MNRAS.487.4025C} {487, 4025}

\bibitem[\protect\citeauthoryear{{Correa Magnus} \& {Vasiliev}}{{Correa Magnus}
  \& {Vasiliev}}{2021}]{Magnus2021}
{Correa Magnus} L.,  {Vasiliev} E.,  2021, arXiv e-prints, \href
  {https://ui.adsabs.harvard.edu/abs/2021arXiv211000018C} {p. arXiv:2110.00018}

\bibitem[\protect\citeauthoryear{{Cunningham} et~al.,}{{Cunningham}
  et~al.}{2020}]{Cunningham2020}
{Cunningham} E.~C.,  et~al., 2020, \mn@doi [\apj] {10.3847/1538-4357/ab9b88},
  \href {https://ui.adsabs.harvard.edu/abs/2020ApJ...898....4C} {898, 4}

\bibitem[\protect\citeauthoryear{{D'Souza} \& {Bell}}{{D'Souza} \&
  {Bell}}{2021}]{DSouza2021}
{D'Souza} R.,  {Bell} E.~F.,  2021, \mn@doi [\mnras] {10.1093/mnras/stab1283},
  \href {https://ui.adsabs.harvard.edu/abs/2021MNRAS.504.5270D} {504, 5270}

\bibitem[\protect\citeauthoryear{{Deason} et~al.,}{{Deason}
  et~al.}{2021}]{Deason2021}
{Deason} A.~J.,  et~al., 2021, \mn@doi [\mnras] {10.1093/mnras/staa3984}, \href
  {https://ui.adsabs.harvard.edu/abs/2021MNRAS.501.5964D} {501, 5964}

\bibitem[\protect\citeauthoryear{{Deg} \& {Widrow}}{{Deg} \&
  {Widrow}}{2013}]{Deg2013}
{Deg} N.,  {Widrow} L.,  2013, \mn@doi [\mnras] {10.1093/mnras/sts089}, \href
  {https://ui.adsabs.harvard.edu/abs/2013MNRAS.428..912D} {428, 912}

\bibitem[\protect\citeauthoryear{{Diemer}, {More}  \& {Kravtsov}}{{Diemer}
  et~al.}{2013}]{Diemer2013}
{Diemer} B.,  {More} S.,   {Kravtsov} A.~V.,  2013, \mn@doi [\apj]
  {10.1088/0004-637X/766/1/25}, \href
  {https://ui.adsabs.harvard.edu/abs/2013ApJ...766...25D} {766, 25}

\bibitem[\protect\citeauthoryear{{Eadie} \& {Juri{\'c}}}{{Eadie} \&
  {Juri{\'c}}}{2019}]{Eadie2019}
{Eadie} G.,  {Juri{\'c}} M.,  2019, \mn@doi [\apj] {10.3847/1538-4357/ab0f97},
  \href {https://ui.adsabs.harvard.edu/abs/2019ApJ...875..159E} {875, 159}

\bibitem[\protect\citeauthoryear{{Erkal} et~al.,}{{Erkal}
  et~al.}{2019}]{Erkal2019}
{Erkal} D.,  et~al., 2019, \mn@doi [\mnras] {10.1093/mnras/stz1371}, \href
  {https://ui.adsabs.harvard.edu/abs/2019MNRAS.487.2685E} {487, 2685}

\bibitem[\protect\citeauthoryear{{Erkal}, {Belokurov}  \& {Parkin}}{{Erkal}
  et~al.}{2020}]{Erkal2020b}
{Erkal} D.,  {Belokurov} V.~A.,   {Parkin} D.~L.,  2020, \mn@doi [\mnras]
  {10.1093/mnras/staa2840}, \href
  {https://ui.adsabs.harvard.edu/abs/2020MNRAS.498.5574E} {498, 5574}

\bibitem[\protect\citeauthoryear{{Erkal} et~al.,}{{Erkal}
  et~al.}{2021}]{Erkal2021}
{Erkal} D.,  et~al., 2021, \mn@doi [\mnras] {10.1093/mnras/stab1828}, \href
  {https://ui.adsabs.harvard.edu/abs/2021MNRAS.506.2677E} {506, 2677}

\bibitem[\protect\citeauthoryear{{Fillingham} et~al.,}{{Fillingham}
  et~al.}{2019}]{Fillingham2019}
{Fillingham} S.~P.,  et~al., 2019, arXiv e-prints, \href
  {https://ui.adsabs.harvard.edu/abs/2019arXiv190604180F} {p. arXiv:1906.04180}

\bibitem[\protect\citeauthoryear{{Fritz}, {Battaglia}, {Pawlowski},
  {Kallivayalil}, {van der Marel}, {Sohn}, {Brook}  \& {Besla}}{{Fritz}
  et~al.}{2018}]{Fritz2018}
{Fritz} T.~K.,  {Battaglia} G.,  {Pawlowski} M.~S.,  {Kallivayalil} N.,  {van
  der Marel} R.,  {Sohn} S.~T.,  {Brook} C.,   {Besla} G.,  2018, \mn@doi
  [\aap] {10.1051/0004-6361/201833343}, \href
  {https://ui.adsabs.harvard.edu/abs/2018A&A...619A.103F} {619, A103}

\bibitem[\protect\citeauthoryear{{Fritz}, {Di Cintio}, {Battaglia}, {Brook}  \&
  {Taibi}}{{Fritz} et~al.}{2020}]{Fritz2020}
{Fritz} T.~K.,  {Di Cintio} A.,  {Battaglia} G.,  {Brook} C.,   {Taibi} S.,
  2020, \mn@doi [\mnras] {10.1093/mnras/staa1040}, \href
  {https://ui.adsabs.harvard.edu/abs/2020MNRAS.494.5178F} {494, 5178}

\bibitem[\protect\citeauthoryear{{Gaia Collaboration} et~al.,}{{Gaia
  Collaboration} et~al.}{2018}]{2018A&A...616A..12G}
{Gaia Collaboration} et~al., 2018, \mn@doi [\aap]
  {10.1051/0004-6361/201832698}, \href
  {https://ui.adsabs.harvard.edu/abs/2018A&A...616A..12G} {616, A12}

\bibitem[\protect\citeauthoryear{{Garavito-Camargo}, {Besla}, {Laporte},
  {Johnston}, {G{\'o}mez}  \& {Watkins}}{{Garavito-Camargo}
  et~al.}{2019}]{Garavito-Camargo2019}
{Garavito-Camargo} N.,  {Besla} G.,  {Laporte} C. F.~P.,  {Johnston} K.~V.,
  {G{\'o}mez} F.~A.,   {Watkins} L.~L.,  2019, \mn@doi [\apj]
  {10.3847/1538-4357/ab32eb}, \href
  {https://ui.adsabs.harvard.edu/abs/2019ApJ...884...51G} {884, 51}

\bibitem[\protect\citeauthoryear{{Garavito-Camargo}, {Besla}, {Laporte},
  {Price-Whelan}, {Cunningham}, {Johnston}, {Weinberg}  \&
  {Gomez}}{{Garavito-Camargo} et~al.}{2020}]{Garavito-Camargo2021}
{Garavito-Camargo} N.,  {Besla} G.,  {Laporte} C. F.~P.,  {Price-Whelan} A.~M.,
   {Cunningham} E.~C.,  {Johnston} K.~V.,  {Weinberg} M.~D.,   {Gomez} F.~A.,
  2020, arXiv e-prints, \href
  {https://ui.adsabs.harvard.edu/abs/2020arXiv201000816G} {p. arXiv:2010.00816}

\bibitem[\protect\citeauthoryear{{Garavito-Camargo}, {Patel}, {Besla},
  {Price-Whelan}, {G{\'o}mez}, {Laporte}  \& {Johnston}}{{Garavito-Camargo}
  et~al.}{2021}]{Garavito-Camargo2021b}
{Garavito-Camargo} N.,  {Patel} E.,  {Besla} G.,  {Price-Whelan} A.~M.,
  {G{\'o}mez} F.~A.,  {Laporte} C. F.~P.,   {Johnston} K.~V.,  2021, \mn@doi
  [\apj] {10.3847/1538-4357/ac2c05}, \href
  {https://ui.adsabs.harvard.edu/abs/2021ApJ...923..140G} {923, 140}

\bibitem[\protect\citeauthoryear{{Garrison-Kimmel}, {Boylan-Kolchin}, {Bullock}
   \& {Lee}}{{Garrison-Kimmel} et~al.}{2014}]{Garrison2014}
{Garrison-Kimmel} S.,  {Boylan-Kolchin} M.,  {Bullock} J.~S.,   {Lee} K.,
  2014, \mn@doi [\mnras] {10.1093/mnras/stt2377}, \href
  {http://adsabs.harvard.edu/abs/2014MNRAS.438.2578G} {438, 2578}

\bibitem[\protect\citeauthoryear{{Gnedin}, {Gould}, {Miralda-Escud{\'e}}  \&
  {Zentner}}{{Gnedin} et~al.}{2005}]{Gnedin2005}
{Gnedin} O.~Y.,  {Gould} A.,  {Miralda-Escud{\'e}} J.,   {Zentner} A.~R.,
  2005, \mn@doi [\apj] {10.1086/496958}, \href
  {https://ui.adsabs.harvard.edu/abs/2005ApJ...634..344G} {634, 344}

\bibitem[\protect\citeauthoryear{{G{\'o}mez}, {Minchev}, {O'Shea}, {Beers},
  {Bullock}  \& {Purcell}}{{G{\'o}mez} et~al.}{2013}]{Gomez2013}
{G{\'o}mez} F.~A.,  {Minchev} I.,  {O'Shea} B.~W.,  {Beers} T.~C.,  {Bullock}
  J.~S.,   {Purcell} C.~W.,  2013, \mn@doi [\mnras] {10.1093/mnras/sts327},
  \href {https://ui.adsabs.harvard.edu/abs/2013MNRAS.429..159G} {429, 159}

\bibitem[\protect\citeauthoryear{{G{\'o}mez}, {Besla}, {Carpintero},
  {Villalobos}, {O'Shea}  \& {Bell}}{{G{\'o}mez} et~al.}{2015}]{Gomez2015}
{G{\'o}mez} F.~A.,  {Besla} G.,  {Carpintero} D.~D.,  {Villalobos} {\'A}.,
  {O'Shea} B.~W.,   {Bell} E.~F.,  2015, \mn@doi [\apj]
  {10.1088/0004-637X/802/2/128}, \href
  {https://ui.adsabs.harvard.edu/abs/2015ApJ...802..128G} {802, 128}

\bibitem[\protect\citeauthoryear{{Hattori} \& {Valluri}}{{Hattori} \&
  {Valluri}}{2020}]{Hattori2020}
{Hattori} K.,  {Valluri} M.,  2020, in {Valluri} M.,  {Sellwood} J.~A.,  eds,
  ~0 Vol. 353, Galactic Dynamics in the Era of Large Surveys. pp 96--100
  (\mn@eprint {arXiv} {1909.03321}), \mn@doi{10.1017/S1743921319008718}

\bibitem[\protect\citeauthoryear{{Hattori}, {Valluri}  \& {Vasiliev}}{{Hattori}
  et~al.}{2021}]{Hattori2021}
{Hattori} K.,  {Valluri} M.,   {Vasiliev} E.,  2021, \mn@doi [\mnras]
  {10.1093/mnras/stab2898}, \href
  {https://ui.adsabs.harvard.edu/abs/2021MNRAS.508.5468H} {508, 5468}

\bibitem[\protect\citeauthoryear{{Helmi}, {Babusiaux}, {Koppelman}, {Massari},
  {Veljanoski}  \& {Brown}}{{Helmi} et~al.}{2018}]{Helmi2018}
{Helmi} A.,  {Babusiaux} C.,  {Koppelman} H.~H.,  {Massari} D.,  {Veljanoski}
  J.,   {Brown} A. G.~A.,  2018, \mn@doi [\nat] {10.1038/s41586-018-0625-x},
  \href {https://ui.adsabs.harvard.edu/abs/2018Natur.563...85H} {563, 85}

\bibitem[\protect\citeauthoryear{{Hernitschek} et~al.,}{{Hernitschek}
  et~al.}{2019}]{Hernitschek2019}
{Hernitschek} N.,  et~al., 2019, \mn@doi [\apj] {10.3847/1538-4357/aaf388},
  \href {https://ui.adsabs.harvard.edu/abs/2019ApJ...871...49H} {871, 49}

\bibitem[\protect\citeauthoryear{{Hernquist}}{{Hernquist}}{1990}]{Hernquist1990}
{Hernquist} L.,  1990, \mn@doi [\apj] {10.1086/168845}, \href
  {https://ui.adsabs.harvard.edu/abs/1990ApJ...356..359H} {356, 359}

\bibitem[\protect\citeauthoryear{{Hoffman}, {Courtois}  \& {Tully}}{{Hoffman}
  et~al.}{2015}]{Hoffman2015}
{Hoffman} Y.,  {Courtois} H.~M.,   {Tully} R.~B.,  2015, \mn@doi [\mnras]
  {10.1093/mnras/stv615}, \href
  {https://ui.adsabs.harvard.edu/abs/2015MNRAS.449.4494H} {449, 4494}

\bibitem[\protect\citeauthoryear{{Hoffman}, {Pomar{\`e}de}, {Tully}  \&
  {Courtois}}{{Hoffman} et~al.}{2017}]{Hoffman2017}
{Hoffman} Y.,  {Pomar{\`e}de} D.,  {Tully} R.~B.,   {Courtois} H.~M.,  2017,
  \mn@doi [Nature Astronomy] {10.1038/s41550-016-0036}, \href
  {https://ui.adsabs.harvard.edu/abs/2017NatAs...1E..36H} {1, 0036}

\bibitem[\protect\citeauthoryear{{Johnston}, {Law}  \& {Majewski}}{{Johnston}
  et~al.}{2005}]{Johnston2005}
{Johnston} K.~V.,  {Law} D.~R.,   {Majewski} S.~R.,  2005, \mn@doi [\apj]
  {10.1086/426777}, \href
  {https://ui.adsabs.harvard.edu/abs/2005ApJ...619..800J} {619, 800}

\bibitem[\protect\citeauthoryear{{Kallivayalil} et~al.,}{{Kallivayalil}
  et~al.}{2018}]{Kallivayalil2018}
{Kallivayalil} N.,  et~al., 2018, \mn@doi [\apj] {10.3847/1538-4357/aadfee},
  \href {https://ui.adsabs.harvard.edu/abs/2018ApJ...867...19K} {867, 19}

\bibitem[\protect\citeauthoryear{{Laporte}, {Johnston}, {G{\'o}mez},
  {Garavito-Camargo}  \& {Besla}}{{Laporte} et~al.}{2018}]{Laporte2018}
{Laporte} C. F.~P.,  {Johnston} K.~V.,  {G{\'o}mez} F.~A.,  {Garavito-Camargo}
  N.,   {Besla} G.,  2018, \mn@doi [\mnras] {10.1093/mnras/sty1574}, \href
  {https://ui.adsabs.harvard.edu/abs/2018MNRAS.481..286L} {481, 286}

\bibitem[\protect\citeauthoryear{{Larson} et~al.,}{{Larson}
  et~al.}{2011}]{Larson2011}
{Larson} D.,  et~al., 2011, \mn@doi [\apjs] {10.1088/0067-0049/192/2/16}, \href
  {https://ui.adsabs.harvard.edu/abs/2011ApJS..192...16L} {192, 16}

\bibitem[\protect\citeauthoryear{{Law} \& {Majewski}}{{Law} \&
  {Majewski}}{2010}]{Law2010}
{Law} D.~R.,  {Majewski} S.~R.,  2010, \mn@doi [\apj]
  {10.1088/0004-637X/714/1/229}, \href
  {https://ui.adsabs.harvard.edu/abs/2010ApJ...714..229L} {714, 229}

\bibitem[\protect\citeauthoryear{{McConnachie} \& {Venn}}{{McConnachie} \&
  {Venn}}{2020}]{McConnachie2020}
{McConnachie} A.~W.,  {Venn} K.~A.,  2020, \mn@doi [\aj]
  {10.3847/1538-3881/aba4ab}, \href
  {https://ui.adsabs.harvard.edu/abs/2020AJ....160..124M} {160, 124}

\bibitem[\protect\citeauthoryear{{Navarro}, {Frenk}  \& {White}}{{Navarro}
  et~al.}{1997}]{NFW1997}
{Navarro} J.~F.,  {Frenk} C.~S.,   {White} S. D.~M.,  1997, \mn@doi [\apj]
  {10.1086/304888}, \href
  {https://ui.adsabs.harvard.edu/abs/1997ApJ...490..493N} {490, 493}

\bibitem[\protect\citeauthoryear{{Nitschai}, {Cappellari}  \&
  {Neumayer}}{{Nitschai} et~al.}{2020}]{Nitschai2020}
{Nitschai} M.~S.,  {Cappellari} M.,   {Neumayer} N.,  2020, \mn@doi [\mnras]
  {10.1093/mnras/staa1128}, \href
  {https://ui.adsabs.harvard.edu/abs/2020MNRAS.494.6001N} {494, 6001}

\bibitem[\protect\citeauthoryear{{Nitschai}, {Eilers}, {Neumayer}, {Cappellari}
   \& {Rix}}{{Nitschai} et~al.}{2021}]{Nitschai2021}
{Nitschai} M.~S.,  {Eilers} A.-C.,  {Neumayer} N.,  {Cappellari} M.,   {Rix}
  H.-W.,  2021, arXiv e-prints, \href
  {https://ui.adsabs.harvard.edu/abs/2021arXiv210605286N} {p. arXiv:2106.05286}

\bibitem[\protect\citeauthoryear{{Nusser} \& {Davis}}{{Nusser} \&
  {Davis}}{2011}]{Nusser2011}
{Nusser} A.,  {Davis} M.,  2011, \mn@doi [\apj] {10.1088/0004-637X/736/2/93},
  \href {https://ui.adsabs.harvard.edu/abs/2011ApJ...736...93N} {736, 93}

\bibitem[\protect\citeauthoryear{{Oluseyi} et~al.,}{{Oluseyi}
  et~al.}{2012}]{Oluseyi2012}
{Oluseyi} H.~M.,  et~al., 2012, \mn@doi [\aj] {10.1088/0004-6256/144/1/9},
  \href {https://ui.adsabs.harvard.edu/abs/2012AJ....144....9O} {144, 9}

\bibitem[\protect\citeauthoryear{{Pardy} et~al.,}{{Pardy}
  et~al.}{2020}]{Pardy2020}
{Pardy} S.~A.,  et~al., 2020, \mn@doi [\mnras] {10.1093/mnras/stz3192}, \href
  {https://ui.adsabs.harvard.edu/abs/2020MNRAS.492.1543P} {492, 1543}

\bibitem[\protect\citeauthoryear{{Patel} et~al.,}{{Patel}
  et~al.}{2020}]{Patel2020}
{Patel} E.,  et~al., 2020, \mn@doi [\apj] {10.3847/1538-4357/ab7b75}, \href
  {https://ui.adsabs.harvard.edu/abs/2020ApJ...893..121P} {893, 121}

\bibitem[\protect\citeauthoryear{{Pe{\~n}arrubia}, {G{\'o}mez}, {Besla},
  {Erkal}  \& {Ma}}{{Pe{\~n}arrubia} et~al.}{2016}]{Penarrubia2016}
{Pe{\~n}arrubia} J.,  {G{\'o}mez} F.~A.,  {Besla} G.,  {Erkal} D.,   {Ma}
  Y.-Z.,  2016, \mn@doi [\mnras] {10.1093/mnrasl/slv160}, \href
  {https://ui.adsabs.harvard.edu/abs/2016MNRAS.456L..54P} {456, L54}

\bibitem[\protect\citeauthoryear{{Petersen} \& {Pe{\~n}arrubia}}{{Petersen} \&
  {Pe{\~n}arrubia}}{2021}]{Petersen2021}
{Petersen} M.~S.,  {Pe{\~n}arrubia} J.,  2021, \mn@doi [Nature Astronomy]
  {10.1038/s41550-020-01254-3}, \href
  {https://ui.adsabs.harvard.edu/abs/2021NatAs...5..251P} {5, 251}

\bibitem[\protect\citeauthoryear{{Petts}, {Read}  \& {Gualandris}}{{Petts}
  et~al.}{2016}]{Petts2016}
{Petts} J.~A.,  {Read} J.~I.,   {Gualandris} A.,  2016, \mn@doi [\mnras]
  {10.1093/mnras/stw2011}, \href
  {https://ui.adsabs.harvard.edu/abs/2016MNRAS.463..858P} {463, 858}

\bibitem[\protect\citeauthoryear{{Posti} \& {Helmi}}{{Posti} \&
  {Helmi}}{2019}]{Posti2019}
{Posti} L.,  {Helmi} A.,  2019, \mn@doi [\aap] {10.1051/0004-6361/201833355},
  \href {https://ui.adsabs.harvard.edu/abs/2019A&A...621A..56P} {621, A56}

\bibitem[\protect\citeauthoryear{{Price-Whelan}, {Hogg}, {Johnston}  \&
  {Hendel}}{{Price-Whelan} et~al.}{2014}]{Price-Whelan2014}
{Price-Whelan} A.~M.,  {Hogg} D.~W.,  {Johnston} K.~V.,   {Hendel} D.,  2014,
  \mn@doi [\apj] {10.1088/0004-637X/794/1/4}, \href
  {https://ui.adsabs.harvard.edu/abs/2014ApJ...794....4P} {794, 4}

\bibitem[\protect\citeauthoryear{{Rocha}, {Peter}  \& {Bullock}}{{Rocha}
  et~al.}{2012}]{Rocha2012}
{Rocha} M.,  {Peter} A. H.~G.,   {Bullock} J.,  2012, \mn@doi [\mnras]
  {10.1111/j.1365-2966.2012.21432.x}, \href
  {https://ui.adsabs.harvard.edu/abs/2012MNRAS.425..231R} {425, 231}

\bibitem[\protect\citeauthoryear{{Ruiz-Lara}, {Gallart}, {Bernard}  \&
  {Cassisi}}{{Ruiz-Lara} et~al.}{2020}]{Ruiz-Lara2020}
{Ruiz-Lara} T.,  {Gallart} C.,  {Bernard} E.~J.,   {Cassisi} S.,  2020, \mn@doi
  [Nature Astronomy] {10.1038/s41550-020-1097-0}, \href
  {https://ui.adsabs.harvard.edu/abs/2020NatAs...4..965R} {4, 965}

\bibitem[\protect\citeauthoryear{{Rusakov}, {Monelli}, {Gallart}, {Fritz},
  {Ruiz-Lara}, {Bernard}  \& {Cassisi}}{{Rusakov} et~al.}{2021}]{Rusakov2021}
{Rusakov} V.,  {Monelli} M.,  {Gallart} C.,  {Fritz} T.~K.,  {Ruiz-Lara} T.,
  {Bernard} E.~J.,   {Cassisi} S.,  2021, \mn@doi [\mnras]
  {10.1093/mnras/stab006}, \href
  {https://ui.adsabs.harvard.edu/abs/2021MNRAS.502..642R} {502, 642}

\bibitem[\protect\citeauthoryear{{Sanders}, {Lilley}, {Vasiliev}, {Evans}  \&
  {Erkal}}{{Sanders} et~al.}{2020}]{Sanders2020}
{Sanders} J.~L.,  {Lilley} E.~J.,  {Vasiliev} E.,  {Evans} N.~W.,   {Erkal} D.,
   2020, \mn@doi [\mnras] {10.1093/mnras/staa3079}, \href
  {https://ui.adsabs.harvard.edu/abs/2020MNRAS.499.4793S} {499, 4793}

\bibitem[\protect\citeauthoryear{{Shipp} et~al.,}{{Shipp}
  et~al.}{2021}]{Shipp2021}
{Shipp} N.,  et~al., 2021, \mn@doi [\apj] {10.3847/1538-4357/ac2e93}, \href
  {https://ui.adsabs.harvard.edu/abs/2021ApJ...923..149S} {923, 149}

\bibitem[\protect\citeauthoryear{{Simon}}{{Simon}}{2018}]{Simon2018}
{Simon} J.~D.,  2018, \mn@doi [\apj] {10.3847/1538-4357/aacdfb}, \href
  {https://ui.adsabs.harvard.edu/abs/2018ApJ...863...89S} {863, 89}

\bibitem[\protect\citeauthoryear{{Vasiliev}, {Belokurov}  \&
  {Erkal}}{{Vasiliev} et~al.}{2021}]{Vasiliev2021}
{Vasiliev} E.,  {Belokurov} V.,   {Erkal} D.,  2021, \mn@doi [\mnras]
  {10.1093/mnras/staa3673}, \href
  {https://ui.adsabs.harvard.edu/abs/2021MNRAS.501.2279V} {501, 2279}

\bibitem[\protect\citeauthoryear{{Wang}, {Mao}, {Zentner}, {Lange}, {van den
  Bosch}  \& {Wechsler}}{{Wang} et~al.}{2020}]{Wang2020}
{Wang} K.,  {Mao} Y.-Y.,  {Zentner} A.~R.,  {Lange} J.~U.,  {van den Bosch}
  F.~C.,   {Wechsler} R.~H.,  2020, \mn@doi [\mnras] {10.1093/mnras/staa2733},
  \href {https://ui.adsabs.harvard.edu/abs/2020MNRAS.498.4450W} {498, 4450}

\bibitem[\protect\citeauthoryear{{Watkins}, {van der Marel}, {Sohn}  \&
  {Evans}}{{Watkins} et~al.}{2019}]{Watkins2019}
{Watkins} L.~L.,  {van der Marel} R.~P.,  {Sohn} S.~T.,   {Evans} N.~W.,  2019,
  \mn@doi [\apj] {10.3847/1538-4357/ab089f}, \href
  {https://ui.adsabs.harvard.edu/abs/2019ApJ...873..118W} {873, 118}

\bibitem[\protect\citeauthoryear{{Wechsler}, {Bullock}, {Primack}, {Kravtsov}
  \& {Dekel}}{{Wechsler} et~al.}{2002}]{Wechsler2002}
{Wechsler} R.~H.,  {Bullock} J.~S.,  {Primack} J.~R.,  {Kravtsov} A.~V.,
  {Dekel} A.,  2002, \mn@doi [\apj] {10.1086/338765}, \href
  {https://ui.adsabs.harvard.edu/abs/2002ApJ...568...52W} {568, 52}

\bibitem[\protect\citeauthoryear{{Weisz}, {Dolphin}, {Skillman}, {Holtzman},
  {Gilbert}, {Dalcanton}  \& {Williams}}{{Weisz} et~al.}{2015}]{Weisz2015}
{Weisz} D.~R.,  {Dolphin} A.~E.,  {Skillman} E.~D.,  {Holtzman} J.,  {Gilbert}
  K.~M.,  {Dalcanton} J.~J.,   {Williams} B.~F.,  2015, \mn@doi [\apj]
  {10.1088/0004-637X/804/2/136}, \href
  {https://ui.adsabs.harvard.edu/abs/2015ApJ...804..136W} {804, 136}

\bibitem[\protect\citeauthoryear{{van den Bosch} \& {Ogiya}}{{van den Bosch} \&
  {Ogiya}}{2018}]{vandenBosch2018b}
{van den Bosch} F.~C.,  {Ogiya} G.,  2018, \mn@doi [\mnras]
  {10.1093/mnras/sty084}, \href
  {http://adsabs.harvard.edu/abs/2018MNRAS.475.4066V} {475, 4066}

\bibitem[\protect\citeauthoryear{{van den Bosch}, {Ogiya}, {Hahn}  \&
  {Burkert}}{{van den Bosch} et~al.}{2018}]{vandenBosch2018a}
{van den Bosch} F.~C.,  {Ogiya} G.,  {Hahn} O.,   {Burkert} A.,  2018, \mn@doi
  [\mnras] {10.1093/mnras/stx2956}, \href
  {http://adsabs.harvard.edu/abs/2018MNRAS.474.3043V} {474, 3043}

\makeatother
\end{thebibliography}

\appendix

\section{Temporal mass-growth resolution}
\label{appendix:timestep}
While backward integrating the orbits using the true mass-growth of the halo, we keep the mass of the central halo constant between snapshots ($\sim250$ Myr). In this section, we test the cost of this step-function assumption. To do this, we first interpolate the mass of the central MW-halo between snapshots using a linear function, and introduce ten steps between every snapshot. That is, the mass is updated ten times between every snapshot. We then backward integrate the orbits of the subhaloes of `iDouglas' to recover its parameters, and calculate the RMS between the recovered parameters and those with the fiducial mass-step resolution. We tabulate the RMS deviations in Table \ref{table:App0} and compare the with the RMS error presented in Table \ref{table:table2}. In majority of the parameters, the RMS deviation due to the assumption of keeping the mass fixed between snapshots is less than 10\% of the RMS error due to the use of parametric models for the potential.

\begin{table}
\label{table:App0}
\caption{The RMS deviations of the recovered parameters of `iDouglas' between a model with temporal mass-growth resolution 10 times smaller than the fiducial model of keeping the mass fixed between snapshots.}
\begin{center}
\begin{tabular}{ c c}
\toprule
code & RMS \\
\midrule
R\_Per1 [kpc] & 1.68\\
R\_Per2 [kpc] & 13.52\\
T\_Per1 [Gyr] & 0.03\\
T\_Per2 [Gyr] & 0.45\\
T\_Per3 [Gyr] & 2.00\\
R\_Apo1 [kpc] & 1.92\\
R\_Apo2 [kpc] & 1.92\\
T\_acc  [Gyr] & 0.49\\
\bottomrule
\end{tabular}
\end{center}
\end{table}

\section{Apocenters are a strong function of binding energy}
\label{appendix:sec1}
\begin{figure}
\includegraphics[trim=0 10 0 10, clip,width=0.9\columnwidth]{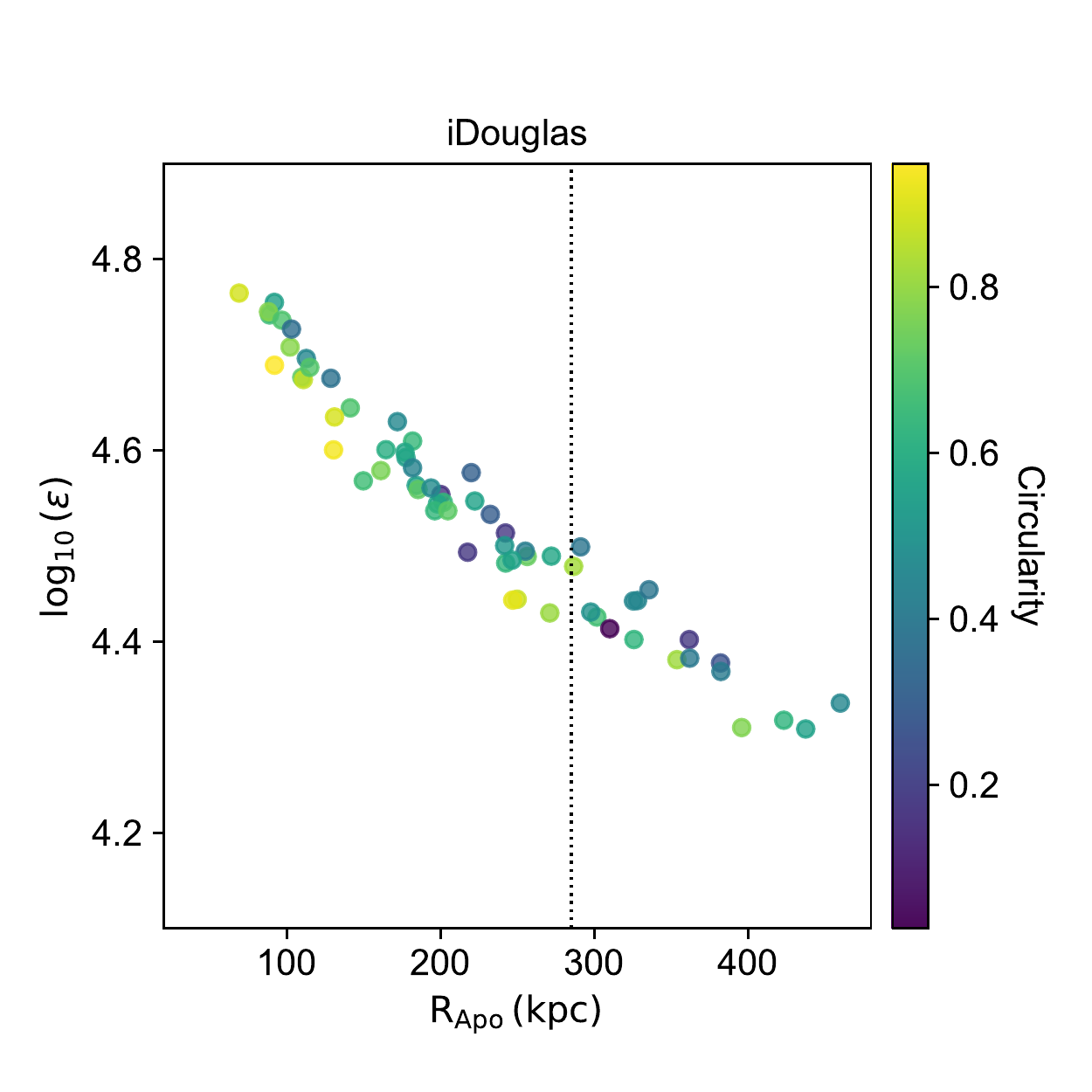}
\includegraphics[trim=0 10 0 10, clip,width=0.9\columnwidth]{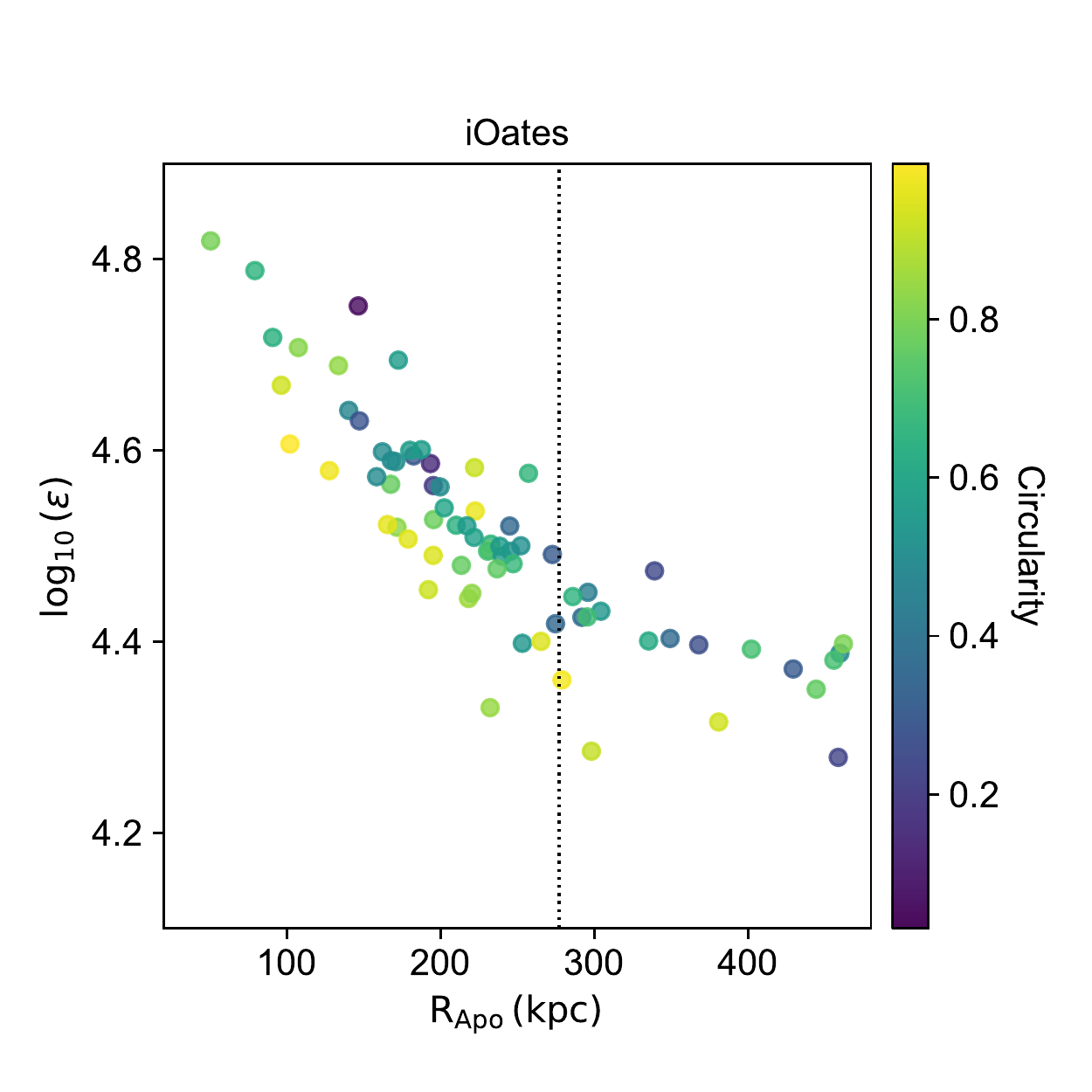}
\caption{The present-day binding energies of the subhaloes as a function of their most recent apocentre for `iDouglas'(top) and `iOates'(bottom). The vertical line marks the virial radius. These plots demonstrate that the apocenter of the orbits of subhaloes are a strong function of the present-day binding energy of the subhaloes.}
\label{fig:App1}
\end{figure}

The apocentre of the orbits of subhaloes are a strong function of their binding energies.  One can intuitively understand this by considering a bound orbit in a one-dimensional effective potential: while the pericentre is a function of both the angular momentum and the potential, the apocentre is a function of primarily the potential as the speed of the satellite at the apocenter is low. We can further see this also in our simulated MW-mass haloes. The most recent apocentres of subhaloes of `iDouglas' are a strong function of their binding energy (top panel of Fig. \ref{fig:App1}). On the other hand, the presence of a massive accretion causes a scattering of binding energies (see Fig. \ref{fig:fig9}). While the apocentres of subhaloes of `iOates' continue to be a function of their binding energy (bottom panel of Fig. \ref{fig:App1}), there is a sizeable scatter in the relation.

\section{Mass Growth of the Halo}
\label{appendix:sec2}
\begin{figure}
\includegraphics[trim=0 30 0 20, clip,width=\columnwidth]{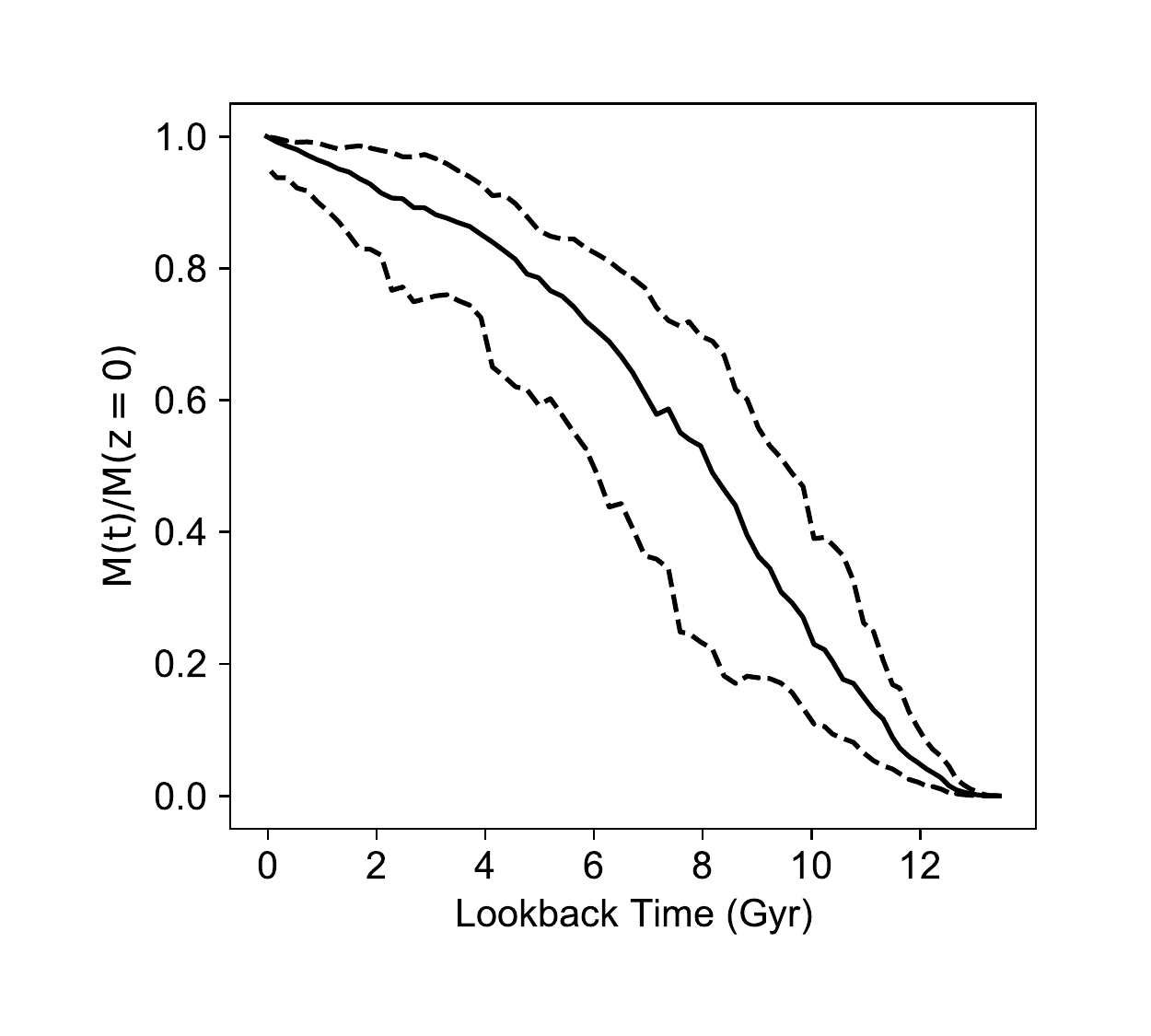}
\caption{The median mass growth of the 48 MW-mass haloes in our sample. The dashed lines show the 10 and 90 percentiles. In the absence of information of the true mass growth of the halo, one can assume that the MW-mass halo grows according to such a median-mass growth-track. This provides an alternative to the assumption of a constant mass commonly used in the literature.}
\label{fig:App2}
\end{figure}

\begin{figure*}
\includegraphics[trim=0 80 0 20, clip,width=0.9\textwidth]{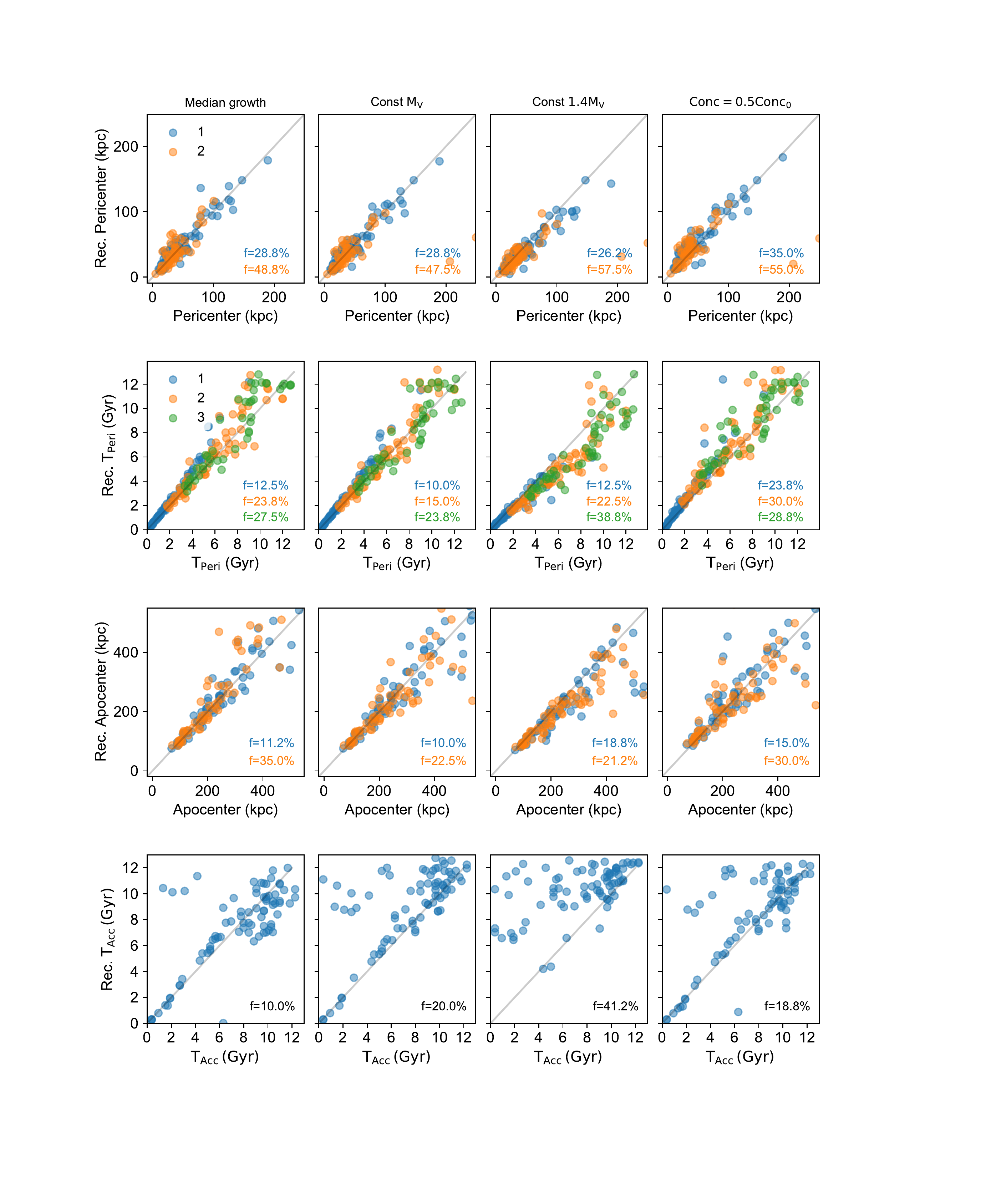}
\caption{Recovery of various parameters of the orbits of the subhaloes of `iDouglas'. Orbits are backward integrated using three models: a) median mass-growth, b) constant mass for the central halo held at its present-day virial masses, c) constant mass for the central halo held at 40\% larger than its present-day virial mass and d) reduced concentration of the main halo ($0.5\mathrm{Conc_{0}}$) while maintaining a constant present-day virial mass. The coloured symbols indicate the first (1), second (2) and third (3) pericentre or apocentre. False positives and false negatives are not show in the figure.}
\label{fig:App3}
\end{figure*}

\begin{figure*}
\includegraphics[trim=0 80 0 10, clip,width=0.9 \textwidth]{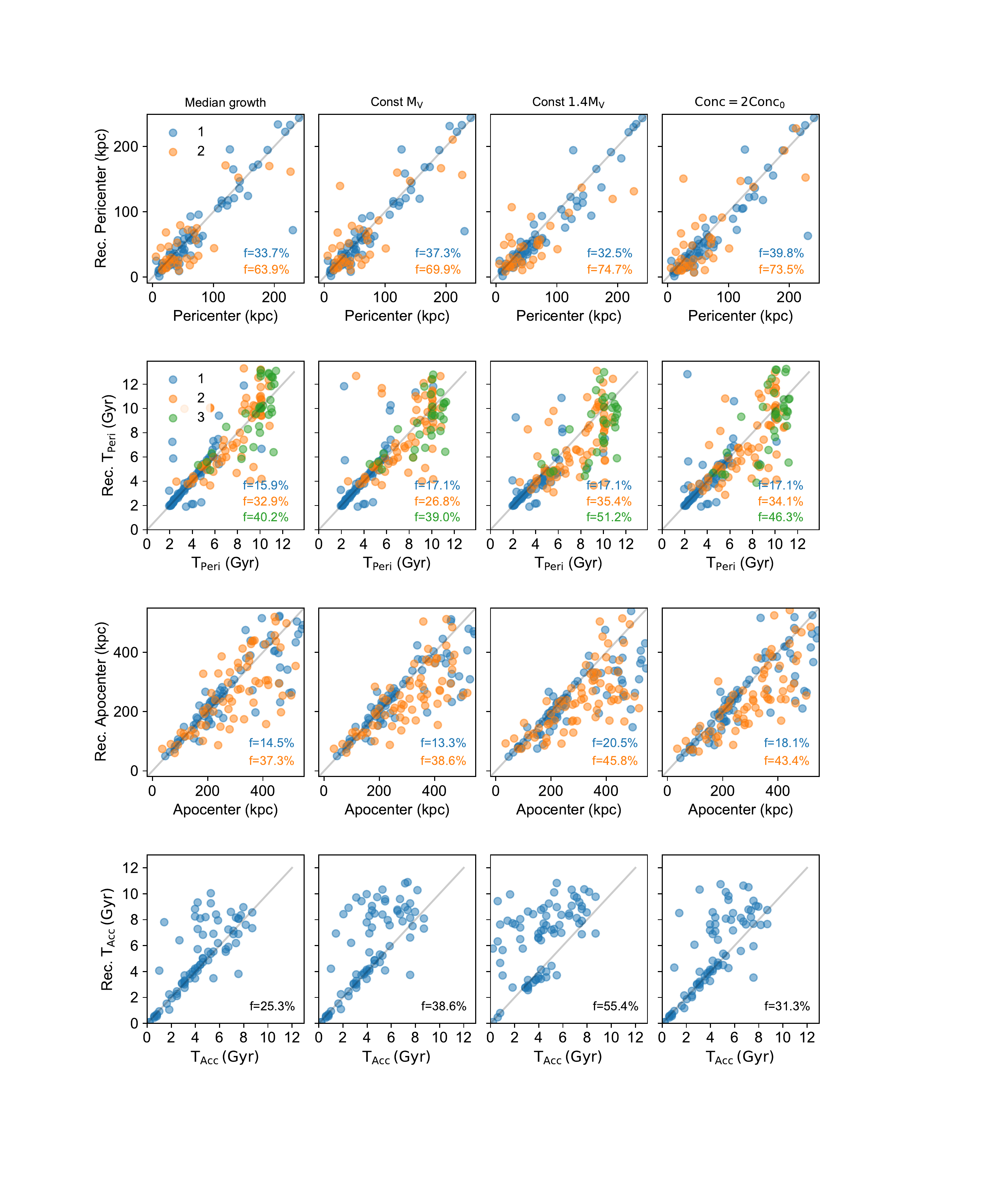}
\caption{Recovery of various parameters of the orbits of the subhaloes of `iOates'. Orbits are backward integrated using three models: a) median mass-growth, b) constant mass for the central halo held at its present-day virial masses, c) constant mass for the central halo held at 40\% larger than its present-day virial mass and d) higher concentration of the main halo ($2\mathrm{Conc_{0}}$) while maintaining a constant present-day virial mass for the main halo and the massive satellite. The coloured symbols indicate the first (1), second (2) and third (3) pericentre or apocentre. False positives and false negatives are not show in the figure.}
\label{fig:App4}
\end{figure*}

Part of the uncertainty associated with the virial mass of the central MW-mass halo is its associated growth history and its mass distribution. In the absence of constraints on the growth history of the MW, it is often convenient to backward integrate the orbits of the dwarf satellites using an assumption of a constant mass with time \citep[e.g.][]{Patel2020}. Alternatively, one can construct a median growth history for MW-mass haloes from simulations (see Fig. \ref{fig:App2}). Moreover, the radial concentration of mass will also play an important role. In this section, we try to understand how the reconstructed orbits of the subhaloes are sensitive to assumptions about the mass-growth and the concentration of the halo. We backward integrate the orbits of the subhaloes of `iDouglas' and `iOates' using four cases: {\it i)} Assume a median mass growth for the central MW-mass halo as shown in Fig. \ref{fig:App2} normalised to the true mass of the halo at $z=0$ while holding the concentartion constant at the value at $z=0$. In the case of `iOates', the recenlty accreted massive satellite is held constant at its maximum mass. {\it ii)} assume that the virial mass of the MW-mass central halo and its concentration are held constant at the values at $z=0$.  In the case of `iOates', the recently accreted massive satellite is held constant at its maximum mass. {\it iii)} Assume that the virial mass for the MW-mass central halo is held constant at 1.4 times the masses at $z=0$ (corresponding to a 40\% increase in the mass of the halo) while the concentration is held constant at the value at $z=0$. In the case of `iOates', the mass of the recently accreted massive satellite is also held constant at 1.4 times its maximum mass. {\it iv)} We hold the concentration of the central MW-mass halo fixed at a value different from its value at $z=0$, while keeping the mass constant at its virial mass. For `iDouglas', we decrease the concentration from 12 to 6. For `iOates', we increase the concentration from 6 to 12. For `iDouglas' which has had a quiescent accretion history, we plot the recovery of various orbital parameters vs.\ their true values in Fig. \ref{fig:App3} and display the statistics in Table D3. For `iOates' which has recently accreted a massive satellite, we plot the recovery of various orbital parameters vs.\ their true values in Fig. \ref{fig:App4} and display the statistics in Table D4.

From these experiments, we notice two additional things, other than what is already mentioned in the main body of the paper. First, in the absence of constraints of the actual mass growth history of the halo, the use of a potential with the median mass growth of haloes or a constant present-day virial mass of the halo are sensible choices --- depending on the problem at hand. If one limits the integration to a few Gyr and has a good handle on the present day virial mass, then using a constant mass has certain advantages, allowing us to recover the first pericentre and apocentre fairly accurately. On the other hand, the assumption of the median mass growth is ideal for deriving the time of accretion of the subhaloes. 

The second thing we notice is that misestimates of the present day virial mass of the central halo leads not only to a larger scatter and outlier fraction, but also to an increase in the bias of many of the derived parameters. In general, larger virial masses of the central halo will lead to orbits circulating for longer lookback time, while for lower virial masses, the subhaloes will escape the gravitational potential of the central halo much earlier. This leads to an overall bias in all the parameters. Hence, attempts to constrain the present-day mass of the MW out to the virial radius are vital to be able to successively backward integrate the orbits of its satellites. 

On the other hand, it appears that misestimates in concentration of the halo do not affect dramatically the scatter or the outlier fraction. In fact, in both the cases where we changed the concentration for `iDouglas' and `iOates', the resulting scatter and outlier fraction is commensurate with those values due to the assumption of a constant mass, suggesting that the changes we have assumed for the concentration do not affect dramatically the backward integration of the orbits. A possible reason for this is that the orbits of the subhaloes are restricted predominately to the outer isothermal part of the halo.

\section{Statistics of recovery of the orbits}
\label{appendix:sec_stat}
In this section, we present various tables containing the statistics of the recovery of backward integrated orbits found in the main text of the paper as well as in Appendix \ref{appendix:sec2}. Table \ref{table:table2} (corresponding to Fig. \ref{fig:fig5}) shows the statistics of the recovery of the orbits for the MW-mass haloes `iDouglas' and `iOates', while Table \ref{table:table3} shows the statistics of the  parameters w.r.t the massive satellite for `iOates', corresponding to Fig. \ref{fig:fig7}. Tables D3 and D4 (corresponding to Figures \ref{fig:App3} \& \ref{fig:App4}) contain the statistics of the parameters of the recovered orbits of `iDouglas' and `iOates' using various assumptions for the mass growth and the concentration of the central MW-mass halo.

Unfortunately, the RMS of the recovered parameters from the backward integrated orbits of `iDouglas' and `iOates' (Table \ref{table:table2}) is extremely sensitive to the false positives/negatives which are assigned the default value of 0, thereby biasing them to large values. In this section, we calculate the constrained Root-Mean-Square (cRMS), i.e., the RMS of the recovered parameters after removing the false positives/negatives - to get a better sense of the RMS in the recovered parameters. We consider the same models considered in Section \ref{sec:backintegrate} and shown in Fig. \ref{fig:fig5}. That is, we integrate the subhaloes of `iDouglas' assuming the true mass growth. For the case of `iOates', we integrate assuming the true mass growth, but also adding the potential of the recently accreted massive satellite. We also backward integrate the subhaloes of `iOates' using only the true mass-growth of the central halo, but neglecting the potential due to the recently accreted massive satellite. We report the cRMS of the recovered parameteres in Table D5, which is smaller than the RMS reported in Table \ref{table:table2}.

\begin{table*}
\begin{center}
\caption{Statistics of the recovery of the orbits for the MW-mass haloes `iDouglas' and `iOates', corresponding to potential models reported in Fig. \ref{fig:fig5}.}
\begin{tabular}{|c@{\hspace{0.3in}}c c c c @{\hspace{0.55in}}c c c c @{\hspace{0.3in}}c c c c}
\toprule
& \multicolumn{4}{c@{\hspace{0.3in}}}{\textbf{iDouglas}} & \multicolumn{8}{c}{\textbf{iOates}} \\[1mm] 
& \multicolumn{4}{c}{\emph{True Growth}} & \multicolumn{4}{c}{\emph{True Growth + Massive Prog.}} & \multicolumn{4}{c}{\emph{True Growth}} \\
code & RMS & bias & scatter & outl. & RMS & bias & scatter & outl. & RMS & bias & scatter & outl.\\
\midrule
R\_Per1 [kpc] & 145.07 & 1.01 & 7.74 & 31.2\% & 123.39 & -6.85 & 38.10 & 36.1\% &180.22 & -1.79 & 22.68 & 47.0\% \\
R\_Per2 [kpc] & 201.31 & -2.06 & 21.59 & 43.8\% & 250.12 & 8.50 & 28.08 & 74.7\% &268.14 & -5.04 & 28.16 & 69.9\% \\
T\_Per1 [Gyr] & 2.69 & -0.30 & 0.39 & 13.8\% & 2.12 & -0.05 & 0.36 & 17.1\% &2.73 & -0.24 & 0.43 & 23.2\% \\
T\_Per2 [Gyr] & 4.28 & -0.41 & 0.82 & 25.0\% & 4.85 & -0.33 & 1.12 & 39.0\% &4.92 & -0.37 & 0.89 & 46.3\% \\
T\_Per3 [Gyr] & 5.27 & -0.49 & 1.09 & 20.0\% & 7.90 & -0.28 & 1.26 & 56.1\% &7.70 & -0.09 & 1.22 & 53.7\% \\
R\_Apo1 [kpc] & 71.84 & -1.78 & 45.23 & 6.2\% & 122.90 & -5.91 & 42.49 & 22.9\% &212.72 & -17.15 & 44.99 & 34.9\% \\
R\_Apo2 [kpc] & 213.62 & -2.08 & 30.68 & 22.5\% & 212.72 & 7.19 & 49.58 & 50.6\% &271.70 & -1.67 & 50.29 & 60.2\% \\
T\_acc  [Gyr] & 2.28 & -0.05 & 1.14 & 11.2\% & 1.76 & -0.12 & 0.62 & 24.1\% &2.16 & -0.17 & 0.66 & 28.9\% \\
\bottomrule
\\
\end{tabular}
\begin{tabular}{ c  @{\hskip 0.3in} c c  @{\hskip 0.55in} c c  @{\hskip 0.3in} c c }
\toprule
& \multicolumn{2}{l}{\textbf{iDouglas}} & \multicolumn{4}{c}{\textbf{iOates}} \\[1mm] 
& \multicolumn{2}{@{\hskip 0.3in}l}{\emph{True Growth}} & \multicolumn{2}{l}{\emph{True Growth + Massive Prog.}} & \multicolumn{2}{c}{\emph{True Growth}}\\
code & False pos. & False neg.  & False pos. & False neg. & False pos. & False neg.\\
\midrule
Per1 & 3.8\% & 1.2\%  & 4.8\% & 0.0\%  & 4.8\% & 3.6\%  \\
Per2 & 10.0\% & 3.8\%  & 14.5\% & 7.2\%  & 10.8\% & 12.0\%  \\
Per3 & 15.0\% & 3.8\%  & 36.6\% & 13.4\%  & 28.0\% & 22.0\%  \\
Apo1 & 0.0\% & 0.0\%  & 0.0\% & 4.8\%  & 0.0\% & 7.2\%  \\
Apo2 & 5.0\% & 6.2\%  & 7.2\% & 12.0\%  & 6.0\% & 21.7\%  \\
\bottomrule
\end{tabular}	
\label{table:table2}
\end{center}
\end{table*}

\begin{table}
\caption{Statistics for parameters w.r.t the massive satellite for `iOates' while modelling the true mass growth of the host and the massive satellite.}	
\label{table:table3}
\centering
\begin{center}
\begin{tabular}{ c c c c c}
\toprule
code & RMS & bias & scatter & outl. \\
\midrule
R\_Per1\_Prog [kpc] & 58.14 & -3.46 & 21.38 & 9.6\%  \\
R\_Per2\_Prog [kpc] & 361.99 & 2.07 & 27.98 & 57.8\%  \\
V\_Per1\_Prog [km/s] & 41.06 & 4.47 & 28.01 & 10.8\%  \\
V\_Per2\_Prog [km/s] & 96.82 & -1.92 & 30.60 & 61.4\%  \\
\bottomrule
\\
\\
\end{tabular}

\begin{tabular}{ c c c}
\toprule
code & False pos. & False neg. \\
\midrule
Per1\_Prog & 2.4\% & 0.0\% \\
Per2\_Prog & 36.1\% & 4.8\% \\
\bottomrule
\end{tabular}
\end{center}
\end{table}

\begin{landscape}
\begin{table}
\label{table:App3}
\centering
\caption{Statistics of the recovery of the orbits for the MW-mass haloes `iDouglas' using four models: a) median mass-growth with fixed concentration b) constant mass and concentration for the central halo held at its present-day virial mass and concentration, c) constant mass for the central halo held at 40\% larger than its present-day virial mass and d) reduced concentration of the main halo ($0.5\mathrm{Conc_{0}}$) while maintaining a constant present-day virial mass.}

\begin{center}
\begin{tabular}{ c c c c c c c c c c c c c c c c c}
	\toprule
& \multicolumn{4}{c}{Median Growth} & \multicolumn{4}{c}{Constant M} & \multicolumn{4}{c}{Constant 1.4 M} & \multicolumn{4}{c}{$0.5\,\mathrm{Conc_{0}}$}  \\
code & RMS & bias & scatter & outl. & RMS & bias & scatter & outl. & RMS & bias & scatter & outl. & RMS & bias & scatter & outl. \\
\midrule
R\_Per1 [kpc] & 139.79 & 1.02 & 8.18 & 28.8\% & 123.81 & 1.57 & 8.59 & 28.8\% & 58.52 & 5.26 & 10.70 & 26.2\% & 184.23 & -1.84 & 9.00 & 35.0\% \\
R\_Per2 [kpc] & 217.85 & -1.75 & 25.80 & 48.8\% & 123.81 & 1.57 & 24.20 & 47.5\% & 208.10 & 4.91 & 8.32 & 57.5\% & 232.36 & -2.15 & 8.38 & 55.0\% \\
T\_Per1 [Gyr] & 2.37 & -0.31 & 0.38 & 12.5\% & 2.28 & -0.32 & 0.44 & 10.0\% & 1.67 & 0.18 & 0.34 & 12.5\% & 2.43 & -0.36 & 0.40 & 23.8\% \\
T\_Per2 [Gyr] & 4.51 & -0.11 & 0.82 & 23.8\% & 3.38 & -0.16 & 0.86 & 15.0\% & 3.41 & 0.96 & 0.85 & 22.5\% & 3.97 & -0.39 & 0.83 & 30.0\% \\
T\_Per3 [Gyr] & 5.58 & -0.30 & 1.06 & 27.5\% & 5.56 & 0.11 & 1.01 & 23.8\% & 6.09 & 1.20 & 1.00 & 38.8\% & 5.21 & -0.50 & 1.02 & 28.8\% \\
R\_Apo1 [kpc] & 91.99 & 3.16 & 42.49 & 11.2\% & 80.93 & 6.86 & 40.80 & 10.0\% & 114.20 & 20.80 & 40.14 & 18.8\% & 151.34 & -3.15 & 44.86 & 15.0\% \\
R\_Apo2 [kpc] & 231.30 & -4.02 & 32.65 & 35.0\% & 209.98 & 7.44 & 39.72 & 22.5\% & 157.64 & 25.96 & 35.06 & 21.2\% & 245.42 & 9.40 & 36.49 & 30.0\% \\
T\_acc  [Gyr] & 2.33 & 0.37 & 1.25 & 10.0\% & 2.91 & -0.46 & 0.92 & 20.0\% & 3.92 & -0.83 & 0.89 & 41.2\% & 2.58 & -0.08 & 0.90 & 18.8\% \\
\bottomrule
\\

\end{tabular}
\begin{tabular}{ c c c c c c c c c}
\toprule
& \multicolumn{2}{c}{Median Growth} & \multicolumn{2}{c}{Constant M} & \multicolumn{2}{c}{Constant 1.4 M} & \multicolumn{2}{c}{$0.5\,\mathrm{Conc_{0}}$} \\
code & False pos. & False neg.  & False pos. & False neg. & False pos. & False neg. & False pos. & False neg.\\
\midrule
Per1 & 3.8\% & 0.0\%  & 3.8\% & 0.0\%  & 3.8\% & 0.0\%  & 3.8\% & 0.0\%\\
Per2 & 11.2\% & 5.0\%  & 8.8\% & 2.5\%  & 12.5\% & 0.0\%  & 10.0\% & 3.8\%\\
Per3 & 15.0\% & 8.8\%  & 16.2\% & 5.0\%  & 27.5\% & 1.2\%  & 13.8\% & 5.0\%\\
Apo1 & 0.0\% & 0.0\%  & 0.0\% & 0.0\%  & 0.0\% & 0.0\%  & 0.0\% & 0.0\%\\
Apo2 & 6.2\% & 10.0\%  & 7.5\% & 7.5\%  & 10.0\% & 1.2\%  & 8.8\% & 6.2\%\\
\bottomrule
\\
\\
\end{tabular}

\caption{Statistics of the recovery of the orbits for the MW-mass haloes `iOates' using four models: a) median mass-growth for the central halo with fixed concentration, constant mass for the satellite held for the massive satellite, b) constant mass for the central halo and the massive satellite held at their present-day virial masses, c) constant mass for the central halo and the massive satellite held at 40\% larger than their present-day virial masses and d) higher concentration of the main halo ($2\mathrm{Conc_{0}}$) while maintaining a constant present-day virial mass for the main halo and the massive satellite.}
\begin{tabular}{ c c c c c c c c c c c c c c c c c}
\toprule
& \multicolumn{4}{c}{Median Growth} & \multicolumn{4}{c}{Constant M} & \multicolumn{4}{c}{Constant 1.4 M} & \multicolumn{4}{c}{$\mathrm{2\,Conc_{0}}$}  \\
code & RMS & bias & scatter & outl. & RMS & bias & scatter & outl. & RMS & bias & scatter & outl. & RMS & bias & scatter & outl. \\
\midrule
R\_Per1 [kpc] & 95.57 & -2.86 & 35.63 & 33.7\% & 128.77 & 0.31 & 14.35 & 37.3\% & 113.57 & 7.93 & 16.79 & 32.5\% & 101.24 & 1.20 & 24.65 & 39.8\% \\
R\_Per2 [kpc] & 216.68 & -10.51 & 48.27 & 63.9\% & 128.77 & 0.31 & 19.96 & 69.9\% & 208.48 & 2.58 & 19.14 & 74.7\% & 212.07 & 6.06 & 40.34 & 73.5\% \\
T\_Per1 [Gyr] & 1.60 & -0.13 & 0.33 & 15.9\% & 2.40 & -0.10 & 0.34 & 17.1\% & 2.21 & 0.30 & 0.56 & 17.1\% & 2.19 & 0.08 & 0.49 & 17.1\% \\
T\_Per2 [Gyr] & 4.38 & -0.09 & 0.96 & 32.9\% & 4.45 & 0.01 & 1.10 & 26.8\% & 4.18 & 0.57 & 1.15 & 35.4\% & 4.26 & 0.27 & 1.17 & 34.1\% \\
T\_Per3 [Gyr] & 6.44 & -0.28 & 1.37 & 40.2\% & 6.93 & 0.39 & 1.48 & 39.0\% & 7.29 & 0.39 & 1.60 & 51.2\% & 6.47 & 0.16 & 1.45 & 46.3\% \\
R\_Apo1 [kpc] & 83.23 & -5.71 & 45.79 & 14.5\% & 74.14 & 2.22 & 43.23 & 13.3\% & 93.29 & 19.89 & 41.13 & 20.5\% & 79.42 & 3.57 & 42.57 & 18.1\% \\
R\_Apo2 [kpc] & 167.34 & -5.47 & 54.49 & 37.3\% & 175.66 & 22.09 & 44.64 & 38.6\% & 162.39 & 23.68 & 37.90 & 45.8\% & 157.54 & 19.17 & 51.34 & 43.4\% \\
T\_acc  [Gyr] & 1.79 & -0.04 & 0.53 & 25.3\% & 2.86 & -0.19 & 0.74 & 38.6\% & 3.50 & -0.12 & 0.90 & 55.4\% & 2.60 & -0.13 & 0.71 & 31.3\% \\
\bottomrule
\\

\end{tabular}
\begin{tabular}{ c c c c c c c c c}
\toprule
& \multicolumn{2}{c}{Median Growth} & \multicolumn{2}{c}{Constant M} & \multicolumn{2}{c}{Constant 1.4 M} & \multicolumn{2}{c}{$\mathrm{2\,Conc_{0}}$} \\
code & False pos. & False neg.  & False pos. & False neg. & False pos. & False neg. & False pos. & False neg.\\
\midrule
Per1 & 3.6\% & 0.0\%  & 4.8\% & 0.0\%  & 4.8\% & 0.0\%  & 3.6\% & 0.0\%\\
Per2 & 14.5\% & 3.6\%  & 10.8\% & 2.4\%  & 13.3\% & 1.2\%  & 13.3\% & 2.4\%\\
Per3 & 30.5\% & 6.1\%  & 25.6\% & 12.2\%  & 35.4\% & 9.8\%  & 29.3\% & 7.3\%\\
Apo1 & 0.0\% & 1.2\%  & 0.0\% & 0.0\%  & 0.0\% & 0.0\%  & 0.0\% & 0.0\%\\
Apo2 & 7.2\% & 6.0\%  & 3.6\% & 7.2\%  & 8.4\% & 2.4\%  & 4.8\% & 4.8\%\\
\bottomrule
\end{tabular}
\end{center}
\end{table}
\end{landscape}

\begin{table}
\label{table:App_RMS}
\caption{The constrained Root-Mean-Square (cRMS) of the recovered parameters of the backward integrated orbits of `iDouglas' and `iOates' corresponding to Table \ref{table:table2}.}
\begin{center}
\begin{tabular}{ c c @{\hspace{0.4in}}c c }
\toprule
& \textbf{iDouglas} & \multicolumn{2}{c}{\textbf{iOates}} \\[1mm] 
code & cRMS &  cRMS  &  cRMS \\
     & \emph{True Growth} & \emph{True Growth}  & \emph{True Growth} \\
	 & & \emph{+ Massive Prog} & \\
\midrule
R\_Per1 [kpc] & 11.31 & 56.52 & 99.79 \\
R\_Per2 [kpc] & 122.09 & 163.59 & 191.58 \\
T\_Per1 [Gyr] & 0.72 & 1.06 & 1.59 \\
T\_Per2 [Gyr] & 1.38 & 2.25 & 2.33 \\
T\_Per3 [Gyr] & 1.25 & 1.90 & 1.65 \\
R\_Apo1 [kpc] & 71.84 & 92.26 & 132.57 \\
R\_Apo2 [kpc] & 95.22 & 118.70 & 148.27 \\
T\_acc  [Gyr] & 2.28 & 1.80 & 2.21 \\
\bottomrule
\end{tabular}
\end{center}
\end{table}

\section{Comparison of sources of error using an alternative metric}
\label{appendix:sec4}
\begin{figure*}
	\includegraphics[trim=0 80 0 30, clip, width=\textwidth,]{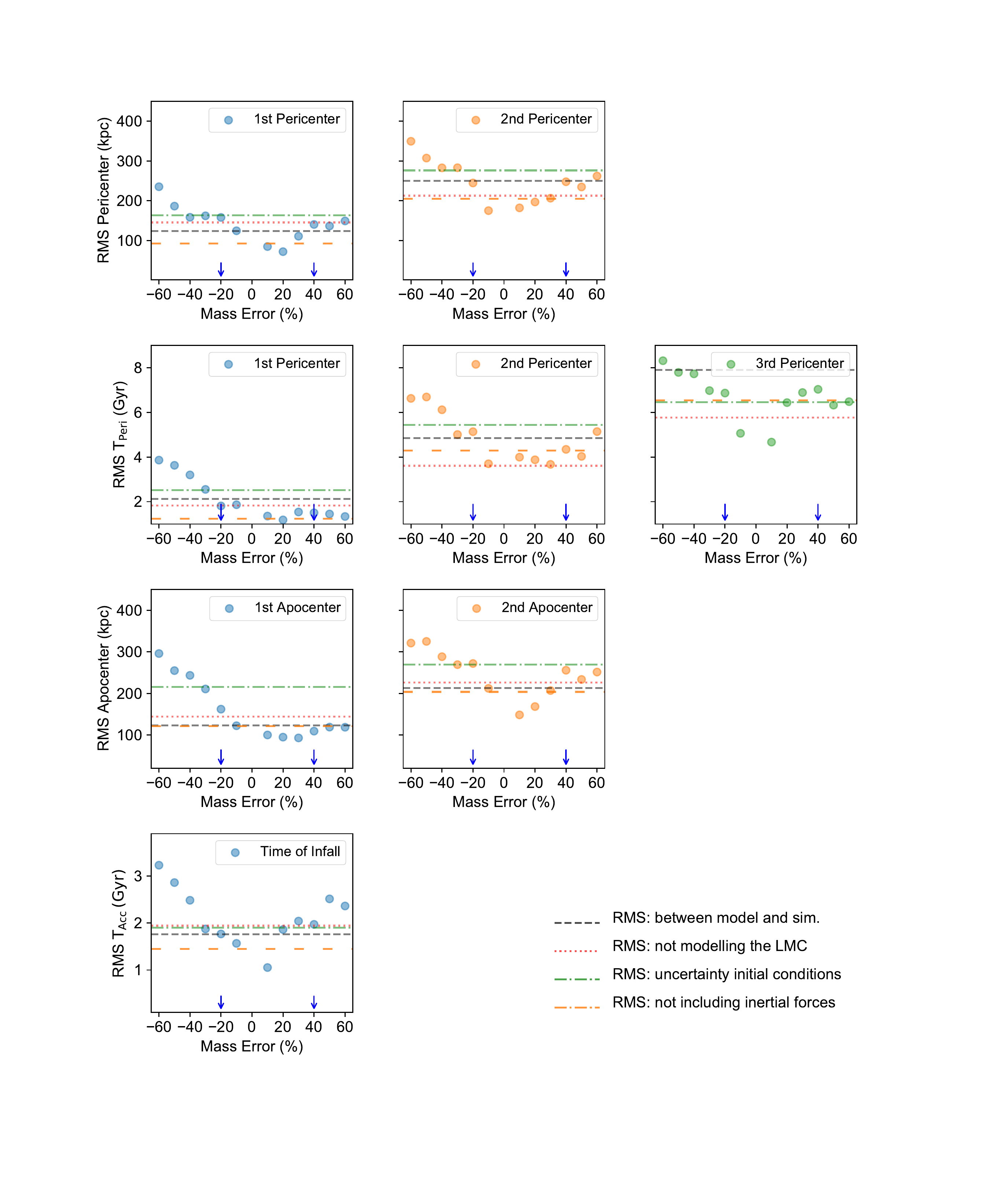}
	\caption{Comparing various sources of error: The RMS of various recovered parameters of `iOates' as a function of changes in the mass of the potential of the central host. The  black dashed horizontal line shows the RMS error between the recovered orbits and the true orbits from the simulation presented in Table \ref{table:table2}. The red dotted horizontal line shows the RMS deviation in the recovered parameters for not modelling the recently accreted massive satellite. The green dot-dashed line shows the RMS deviation due to uncertainties in the initial conditions. We suggest that the RMS deviation due to a -20\% to 40\% uncertainty in the mass of the central MW halo is comparable to the RMS error due to the use of the parametric models of the potential. We indicate using arrows the suggested range.}
	\label{fig:App5}
\end{figure*}

In Section \ref{sec:compare}, we compare various sources of error using the `outlier fraction' based on the equation defined in Section \ref{subsec:parameter}, which measures only the fraction of subhaloes whose deviations in the recovered parameters are generally greater than 30\%. While having many advantages especially with dealing with false positives/negatives, it does not give us a sense of the scatter of the parameters recovered successfully. Another possible metric one can use is the RMS. While such a metric can measure the scatter, it is also very sensitive to false positives/negatives in the recovered parameters and thus the extreme tails of the distribution. In Fig. \ref{fig:App5}, we present the RMS deviations of the parameters of the recovered orbits by varying the mass of the potential from those obtained using the true halo mass, and compare them to the RMS errors due to the use of the parametric forms of the potential. We also compare them to the RMS deviations due to not modelling the recently accreted massive satellite as well as the uncertainty in the initial conditions. Similar to Fig. \ref{fig:fig8}, we find that RMS deviations are asymmetric around positive and negative changes in the mass of the central potential. We also find that the RMS errors of the recovered parameters due to the use of parametric forms of the potential are comparable to the RMS deviations caused by a 20\% decrease or a $\sim40-50\%$ increase in the mass of the potential. Moreover, we also find that the RMS deviations due to not modelling the recently accreted massive satellite are comparable to the RMS errors due to the use of parametric forms of the potential. The `distance of the first apocentre' is the only parameter which stands out with the RMS deviations being double the RMS errors. The lack of modelling the recently accreted massive satellite particularly particularly affects the depth of the potential during the time of its infall into the central host. This in turn affects the most recent apocentric distance. Finally, even the RMS deviations due to the uncertainties in the initial conditions are comparable to RMS errors due to the use of parametric forms of the potential. However, with newer data, we expect that the uncertainties in the initial conditions will decrease over time. We conclude that these tests with the RMS metric broadly reflect what we found with the outlier fraction in Fig. \ref{fig:fig8}, and that the errors caused by the contribution to the total error budget due the use of the parametric forms of the potential rival i) a 30\% uncertainty in the virial mass of the MW as well as ii) not modelling the potential of the recently accreted massive satellite. 

% Don't change these lines
\bsp	% typesetting comment
\label{lastpage}
\end{document}